%% file: p.tex
\documentclass[letterpaper,twocolumn,10pt]{article}
\usepackage{usenix}

\AtBeginDocument{%
  }

\input{pkgs}

\newcommand{\Latlantis}{\textsc{Atlantis}\xspace} 
\newcommand{\Lbuttercup}{\textsc{Buttercup}\xspace} 
\newcommand{\Lroboduck}{\textsc{RoboDuck}\xspace} 
\newcommand{\Lfuzzingbrain}{\textsc{FuzzingBrain}\xspace}
\newcommand{\Lartiphishell}{\textsc{Artiphishell}\xspace} 
\newcommand{\Lbugbuster}{\textsc{BugBuster}\xspace} 
\newcommand{\Llacrosse}{\textsc{Lacrosse}\xspace}

\newcommand{\atlantis}{\texttt{AT}\xspace}
\newcommand{\buttercup}{\texttt{TB}\xspace}
\newcommand{\roboduck}{\texttt{TI}\xspace}
\newcommand{\fuzzingbrain}{\texttt{FB}\xspace}
\newcommand{\artiphishell}{\texttt{SP}\xspace}
\newcommand{\bugbuster}{\texttt{42}\xspace}
\newcommand{\lacrosse}{\texttt{LC}\xspace}

\input{cmds}
\input{rev}

\usepackage[available]{usenixbadges}

\begin{document}

\input{hdr}

\maketitle
\ifdefined\DRAFT\else\thispagestyle{empty}\fi 

\begingroup
\renewcommand{\thefootnote}{\textsection}%
\footnotetext{Work primarily conducted at Georgia Institute of Technology.}%
\endgroup

\input{abstract}

\sloppy
\input{intro}

\input{bg}

\input{comp}
\input{cp}
\input{crs-intro}
\input{crs-tech}

\input{ana}

\input{lessons}
\input{validity}
\input{conclusion}
\input{ack}

\input{ethics}
\input{openscience}

\bibliographystyle{plainurl}
\bibliography{p,sslab,conf}

\appendix
\input{app}

\end{document}

%% file: pkgs.tex
\hypersetup{citecolor=blue,linkcolor=blue,breaklinks=true,colorlinks=true}
\usepackage{xurl} 
\usepackage{amsmath,amsopn,amssymb}
\usepackage{subcaption}
\usepackage{endnotes,microtype,xspace,graphicx,fancyvrb,multirow}
\usepackage{booktabs}
\usepackage{tabularx}
\usepackage[table]{xcolor} 
\usepackage{longtable}
\usepackage{tabularray}
\UseTblrLibrary{booktabs}
\usepackage{adjustbox}
\usepackage{makecell}
\usepackage{threeparttable}
\usepackage{pifont}
\usepackage{wasysym}
\usepackage{comment}
\usepackage{array,underscore,relsize}
\usepackage[super]{nth}
\usepackage[T1]{fontenc}
\usepackage{pmboxdraw}
\usepackage[normalem]{ulem}
\usepackage[inline]{enumitem}

\usepackage{fp}
\usepackage{siunitx}

\usepackage{minted}

\usepackage{balance}

\usepackage{tikz}
\usetikzlibrary{positioning,fit}
\usepackage{pgfplots}
\pgfplotsset{compat=1.17}

%% file: cmds.tex
\newcommand{\fullcirc}{\raisebox{0.25ex}{\scalebox{0.7}{\CIRCLE}}}
\newcommand{\emptycirc}{\raisebox{0.25ex}{\scalebox{0.8}{\Circle}}}

\newcommand{\cc}[1]{\mbox{\smaller[0.5]\texttt{#1}}}

\fvset{fontsize=\scriptsize,xleftmargin=8pt,numbers=left,numbersep=5pt}

\IfFileExists{code/fmt.tex}{\input{code/fmt}}{}

\def\Snospace~{\S{}}

\newif\ifdraft\drafttrue
\newif\ifnotes\notestrue
\ifdraft\else\notesfalse\fi

\input{glyphtounicode}
\newcolumntype{R}[1]{>{\raggedleft\let\newline\\\arraybackslash\hspace{0pt}}p{#1}}

\newcommand{\squishlist}{
\begin{itemize}
}
\newcommand{\squishend}{
  \end{itemize}
}

\usepackage{tikz}

\usepackage{xstring}
\newcommand{\PP}[1]{
\noindent{\bf \IfEndWith{#1}{.}{#1}{#1.}}
}

\newcommand{\BB}[1]{\emph{\IfEndWith{#1}{.}{#1}{#1.}}}

\newcommand{\V}{\checkmark}
\newcommand{\tsp}{\hspace*{0.5em}} 
\newcommand{\tmid}{{\footnotesize\textSFviii\textSFx}} 
\newcommand{\tlast}{{\footnotesize\textSFii\textSFx}}  
\newcommand{\tcont}{{\footnotesize\textSFxi\phantom{\textSFx}}} 
\newcommand{\X}{{\footnotesize $\times$}\xspace}

\newcommand{\cpvd}[2]{\cc{#1}{\smaller$\blacktriangle$}\cc{#2}}
\newcommand{\cpvf}[2]{\cc{#1}{\smaller$\square$}\cc{#2}}

\newcommand{\eg}{\textit{e}.\textit{g}.}

\newcommand{\boxbeg}{%
\vspace{2px}%
\noindent\begin{tabular}{|l|}\hline
\begin{minipage}{\columnwidth}%
\vspace{2px}%
\noindent
}

\newcommand{\boxend}{%
\vspace{2px}%
\end{minipage}\\ \hline
\end{tabular}%
\vspace{-10pt}%
}

\newcommand{\findingbox}[2]{%
  \vspace{4pt}
  \noindent
  \begin{tikzpicture}
    \node[inner sep=6pt, text width=0.96\columnwidth, fill=gray!10, align=justify] (box) {%
      \textbf{#1}~#2%
    };
    \fill[blue!70] (box.north west) rectangle ([xshift=3pt]box.south west);
  \end{tikzpicture}%
}

\definecolor{ForestGreen}{RGB}{34,139,34}

\newcommand{\TeamScoreFormula}{%
\[
S_{\text{Team}} = \sum S_{\text{Challenge}}
\]
}

\newcommand{\ChallengeScoreFormula}{%
\[
S_{\text{Challenge}} = AM \times
(S_{\text{PoV}} + S_{\text{Patch}} + S_{\text{SARIF}} + S_{\text{Bundle}})
\]
}

\newcommand{\BundleScoreFormula}{%
\[
S_{\text{Bundle}} =
\pm \Big(
\underbrace{0.5 (S_{\text{PoV}} + S_{\text{Patch}})}_{\text{PoV-Patch}}
+ \underbrace{\,1\,}_{\text{PoV-SARIF}}
+ \underbrace{\,2\,}_{\text{Patch-SARIF}}
\Big)
\]
}

\newcommand{\AccuracyMultiplierFormula}{%
\begin{center}
\begin{minipage}[c]{0.8\columnwidth}
\[
AM = 1 - (1 - r)^4, \quad r = \frac{n_{\text{acc}}}{n_{\text{acc}} + n_{\text{inacc}}}
\]
\end{minipage}%
\begin{minipage}[c]{0.2\columnwidth}
\centering
\phantom{\scriptsize 0.5}\\[-0em]
\begin{tikzpicture}
\begin{axis}[
    width=2.8cm, height=2.4cm,
    tick label style={font=\scriptsize},
    xmin=0, xmax=1, ymin=0, ymax=1.1,
    xtick={0,0.5,1}, ytick={0,0.5,1},
    grid=major, grid style={dashed, gray!30},
]
\addplot[blue, thick, domain=0:1, samples=50] {1 - (1-x)^4};
\end{axis}
\end{tikzpicture}
\end{minipage}
\end{center}
}

\newcommand{\TimeDecayFormula}{%
\begin{center}
\begin{minipage}[c]{0.8\columnwidth}
\[
\mathit{Score} = \mathit{weight} \times \tau, \quad
\tau = 0.5 + \frac{\text{remaining time}}{2 \times \text{total duration}}
\]
\end{minipage}%
\begin{minipage}[c]{0.2\columnwidth}
\centering
\phantom{\scriptsize 0.5}\\[-0.5em]
\begin{tikzpicture}
\begin{axis}[
    width=2.8cm, height=2.4cm,
    tick label style={font=\scriptsize},
    xmin=0, xmax=8, ymin=0, ymax=1.1,
    xtick=\empty, ytick={0,0.5,1},
    grid=major, grid style={dashed, gray!30},
]
\addplot[blue, thick, domain=0:6, samples=2] {1 - x/12};
\addplot[blue, thick] coordinates {(6,0.5) (6,0)};
\addplot[blue, thick, domain=6:8, samples=2] {0};
\end{axis}
\end{tikzpicture}
\end{minipage}
\end{center}
}

%% file: rev.tex
\gdef\therev{13f5035}
\gdef\thedate{2026-06-12 18:33:29 -0400}

%% file: hdr.tex
\date{}

\title{\Large \bf SoK: DARPA's AI Cyber Challenge (AIxCC):\\Competition Design, Architectures, and Lessons Learned}

\ifdefined\DRAFT
 \pagestyle{fancyplain}
 \lhead{Rev.~\therev}
 \rhead{\thedate}
 \cfoot{\thepage\ of \pageref{LastPage}}
\else
 \pagestyle{empty}
\fi

\author{
Cen Zhang$^{\dagger\ast\mathsection}$\;
Younggi Park$^\natural$\;
Fabian Fleischer$^\dagger$\;
Yu-Fu Fu$^\dagger$\;
Jiho Kim$^\dagger$\;
Dongkwan Kim$^{\ast\mathsection}$\;
\\
Youngjoon Kim$^\dagger$\;
Qingxiao Xu$^\ddagger$\;
Andrew Chin$^\dagger$\;
Ze Sheng$^\ddagger$\;
Hanqing Zhao$^\dagger$\;
\\
Michael Pelican$^\P$\;
David J.\ Musliner$^\P$\;
Jeff Huang$^\ddagger$\;
Jon Silliman$^\|$\;
Mikel Mcdaniel$^\|$\;
\\
Jefferson Casavant$^\|$\;
Isaac Goldthwaite$^\|$\;
Nicholas Vidovich$^\|$\;
Matthew Lehman$^\|$\;
Taesoo Kim$^{\dagger\ast}$\;
\\\\
\emph{$^\dagger$ Georgia Institute of Technology},
\emph{$^\ddagger$ Texas A\&M University},
\emph{$^\natural$ Independent Researcher},
\\
\emph{$^\P$ Smart Information Flow Technologies (SIFT)},
\emph{$^\|$ Kudu Dynamics},
\emph{$^\ast$ Microsoft}
}

%% file: abstract.tex
\begin{abstract}
DARPA's AI Cyber Challenge (AIxCC, 2023--2025)
is the largest competition to date
for building fully autonomous Cyber Reasoning Systems (CRSs)
that leverage recent advances in AI---particularly large language models (LLMs)---to
discover and remediate vulnerabilities in real-world open-source software.
This paper presents the first systematic analysis of AIxCC.
Drawing on design documents, source code,
execution traces, and discussions
with organizers and all finalist teams,
we examine the competition's structure and key design decisions,
characterize the architectural approaches of finalist CRSs,
and analyze competition results beyond the final scoreboard.
Our analysis reveals the factors that truly drove CRS performance,
identifies genuine technical advances achieved by teams,
and exposes limitations that remain open for future research.
We conclude with lessons for organizing future competitions
and broader insights toward deploying autonomous CRSs in practice.
\end{abstract}

%% file: intro.tex
\section{Introduction}
\label{s:intro}

Open-source software (OSS) underpins critical infrastructure,
yet scaling vulnerability discovery and remediation remains challenging.
DARPA's AI Cyber Challenge (AIxCC, 2023--2025)
addresses this by challenging teams
to build fully autonomous Cyber Reasoning Systems (CRSs)
that leverage large language models (LLMs)
to discover and patch vulnerabilities in real-world C and Java projects.
The final competition in August 2025
represents the largest-scale evaluation of autonomous vulnerability analysis to date:
around 143 hours of fully autonomous operation,
CRSs from seven finalist teams analyzed 53 challenge projects
derived from critical infrastructure software,
each equipped with \$85K in cloud compute and \$50K in LLM API credits.

Despite the competition's completion,
no systematic study has examined AIxCC's design rationale,
the technical approaches employed by participating teams,
or the lessons that emerged from this large-scale competition.
Such an analysis would benefit multiple communities:
competition organizers designing future challenges,
security researchers developing advanced vulnerability detection and patching techniques,
and practitioners seeking AI-driven security solutions.

To fill this gap,
we conducted a systematic study of the final competition,
drawing on multiple primary sources:
all seven finalist CRS codebases and whitepapers,
the complete competition database
(challenges, results, and execution traces) from organizers,
and discussions with organizers and all finalist teams.
Our analysis examines AIxCC from three perspectives:
the design decisions that shaped the competition,
the architectural and technical choices made by finalist teams,
and competition results and their implications.
Specifically, we address the following research questions:

\squishlist
	\item \textbf{RQ1:} How is AIxCC designed
	      to guide and evaluate AI-driven vulnerability discovery and patching?
	\item \textbf{RQ2:} What architectural and technical approaches
	      did finalist teams employ?
	\item \textbf{RQ3:} What insights emerge from the results?
	\item \textbf{RQ4:} What are the lessons and future directions?
\squishend

Our work makes the following contributions:
\squishlist
\item A systematic analysis of AIxCC's competition design,
      covering design and scoring principles,
      challenge construction, and execution guidelines.
\item A taxonomy of CRS architectures and techniques across all seven finalist teams,
      spanning vulnerability discovery, patching, report triage, and bundling.
\item In-depth result analysis
      that reveals the true factors behind CRS performance,
      with per-vulnerability analysis for technical insights.
\item Lessons on translating competition outcomes
      to industry deployment and research,
      plus future directions.
\squishend
All data and artifacts will be released publicly upon acceptance
(some subject to DARPA's timeline; see \autoref{s:openscience}).

%% file: bg.tex
\section{Background: AIxCC as Competition}
\label{s:bg}

\PP{AIxCC}
AIxCC~\cite{aixcc} (2023--2025) is a DARPA/ARPA-H competition
to advance fully autonomous vulnerability discovery and remediation
for open-source software,
in collaboration with AI providers (Anthropic, Google, Microsoft, OpenAI),
Linux Foundation, OpenSSF, Black Hat USA, and DEF CON.
From 42 entrants, seven teams advanced through the semifinal
(ASC, DEF CON 2024)~\cite{aixcc-semifinal}
to the final (AFC, DEF CON 2025), the focus of this paper.\footnote{
This artifact represents the authors' own statements and does not constitute an official DARPA statement.}
The final ran for around 143 hours across seven phases (P1--P7),
during which the finalists deployed CRSs autonomously to analyze
53 challenge projects (CPs) in C and Java
under \$85,000 in Azure compute
and \$50,000 in LLM API credits per team.
All finalist CRSs, challenges, and infrastructure
are being open-sourced~\cite{aixcc-archive}.

\PP{Comparison with CGC}
AIxCC is a spiritual successor to DARPA's Cyber Grand Challenge
(CGC, 2014--2016)~\cite{cgc} after a near-decade gap;
each marks a pivotal inflection point for fully autonomous CRSs.
The two differ in two ways.
\ding{192} \emph{Scope}: CGC focused on binary exploitation on DECREE OS~\cite{cgc-walker},
while AIxCC targets vulnerability discovery and remediation
for real-world OSS in C and Java.
\ding{193} \emph{AI emphasis}: AIxCC provides LLM infrastructure from AI providers,
highlighting LLM-based techniques unavailable during CGC.

\PP{Acronyms}
\autoref{s:app:abbr} provides the mappings and rules
for abbreviations used in this paper.

\PP{Our Methodology}
\ding{192} \emph{Competition design} (\autoref{s:comp}):
we extract the design rationale from the organizers' reflection documents
and whitepapers,
refined through online discussions and confirmed by them.
\ding{193} \emph{CRS technique taxonomy} (\autoref{s:crs:intro}, \autoref{s:crs}):
each team's profile is grounded primarily in the enabled functionalities
of its submission-version code;
at least two security experts independently reviewed each codebase
and cross-validated the resulting profile.
Team whitepapers, blogs, and a structured questionnaire
serve as supplementary sources.
The questionnaire covers common competition reflections and team-specific technical designs;
one team responded by email and the others through dedicated meetings.
\ding{194} \emph{Outcome analysis} (\autoref{s:analysis}):
per-team outcomes are derived from the official competition database
(logs, traces, scores, etc.), challenge code, and vulnerability data;
patch outcome labels (\autoref{s:analysis:annotation})
are additionally cross-validated by at least two security experts,
as with the taxonomy.

%% file: comp.tex
\begin{figure}[t]
\centering
\includegraphics[width=\columnwidth]{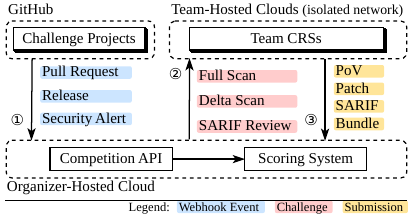}
\caption{Final competition workflow:
\ding{192}~triggered by webhook, \ding{193}~challenge dispatched, \ding{194}~result submitted.}
\label{f:infra}
\end{figure}

\section{Competition Design}
\label{s:comp}

AIxCC grounds its CRS evaluation in real-world OSS development workflows:
the final competition embeds in GitHub,
with each CRS capability triggered by an actual development event.
\autoref{f:infra} shows the workflow.
The four CRS capabilities each address a real development moment:
\squishlist
\item \emph{Full Scan}:
    when developers tag a new release,
    detect and patch vulnerabilities across the full codebase.
\item \emph{Delta Scan}:
    when new code is merged via pull requests,
    conduct targeted analysis on the incremental changes.
\item \emph{SARIF Review}:
    when developers receive static-analyzer security alerts in SARIF\footnote{SARIF (Static Analysis Results Interchange Format)~\cite{sarif-oasis} is a JSON format for static analysis tool output; each report identifies a potential vulnerability with file path, code region, description, and CWE classification.} format,
    classify each as valid (a real vulnerability) or invalid.
\item \emph{Report Synthesis}:
    after analysis,
    correlate all above findings
    into per-vulnerability reports.
\squishend

Organizers send two kinds of challenges (Full Scan or Delta Scan),
some accompanied by SARIF broadcasts for review.
A CRS may submit up to four kinds of evidence per challenge:
\squishlist
\item \emph{Proof of Vulnerability (PoV)} ($[1, 2]$ pts):
    an input that triggers abnormal execution (\eg, a crash).
\item \emph{Patch} ($[3, 6]$ pts):
    a fix that resolves the vulnerability while preserving functionality.
\item \emph{SARIF assessment} ($[0.5, 1]$ pt):
    a judgment on whether a SARIF report is valid.
\item \emph{Bundle} ($[-7, 7]$ pts):
    a grouping that links related findings for the same vulnerability.
\squishend

\PP{Scoring: Defining a ``Good CRS''}
Each submission's weight reflects how much developer time and effort it saves
(or wastes, when wrong):
\ding{192} Patch scores high since it provides a direct solution for the security issue;
\ding{193} PoV scores lower since it only proves the bug exists;
\ding{194} SARIF scores lowest since it judges a statically-identified security issue;
\ding{195} Bundle scores both extremes,
up to +7 for fully correct linkage or a penalty for any incorrect,
since correct linkage saves the most triage effort
and wrong linkage wastes the most.

Two further mechanisms reward fast and correct submissions.
\emph{Time-decay} grants full points for immediate submissions
and half for last-minute ones.
The \emph{accuracy multiplier} penalizes a CRS's per-challenge total
by its accuracy rate on that challenge,
with a non-linear shape that balances technique exploration with practicality:
high accuracy is barely affected
(\eg, 90\% $\rightarrow$ negligible penalty),
while low accuracy is steeply penalized
(50\% $\rightarrow$ 6\% reduction; 40\% $\rightarrow$ 13\% reduction).
See \cite{sok-website} for full scoring details.

\PP{Execution: Surfacing Problems Early}
AIxCC stretched across two years and 100+ collaborating organizations,
in a field where AI capability evolved rapidly.
Perfecting the design in advance was unrealistic,
so organizers iterated at two scales
to surface and fix issues before the scored competition.

\emph{From semifinal to final.}
The semifinal narrowed 42 teams to 7 finalists
while exposing two design flaws that the final corrected:
\ding{192} the semifinal's self-PoV requirement disadvantaged CRSs
strong at patching but weak at discovery;
the final scored patches against all PoVs from all teams.
\ding{193} Per-repository challenge adoption demanded significant engineering;
the final adopted OSS-Fuzz instead.
With fewer teams to support,
the final granted broader team autonomy
(per-team budgets and self-managed Azure infrastructure)
and dedicated pre-competition support,
in place of the semifinal's constrained sandbox.

\emph{Within the final.}
Beyond internal infrastructure testing,
the final ran exhibition rounds with the finalist teams.
Three unscored rounds used a separate challenge set that mirrored the scored setting,
letting both teams and organizers iterate on their systems and processes.

%% file: cp.tex
\section{Challenge Projects}
\label{s:cp}

\input{tbl/cp_details/overview_table}

The final comprised 48 scored challenge projects (CPs)
drawn from 24 OSS-Fuzz~\cite{oss-fuzz} repositories,
together containing 63 challenge project vulnerabilities (CPVs) (\autoref{t:cp-overview}).
Five additional CPs lacked fuzzing harnesses and were excluded from scoring.
Organizers selected critical-infrastructure and healthcare-critical OSS,
onboarding new projects to OSS-Fuzz where needed.
To avoid AI training contamination,
most CPVs are hand-crafted synthetics inspired by historical N-day issues,
with a few genuine 0-days surfaced during challenge development.

\PP{Repositories}
The 24 repositories (14~C and 10~Java) span
diverse application categories,
from image-processing libraries
to healthcare-critical software.
The number of harnesses per repository varies considerably,
from a single harness (\cc{dav1d}, \cc{libavif}, etc.) to 55 (\cc{ndpi}),
and repository sizes range
from 16K (\cc{libexif}) to 4.9M (\cc{wireshark}) lines of code.

\PP{Challenge Projects}
The 48 CPs comprise 16 full-mode (vulnerabilities anywhere in the repository)
and 32 delta-mode (vulnerabilities within a PR diff) challenges,
of which eight contain no injected CPV
and serve solely as 0-day discovery targets.
For delta challenges,
CPV-containing delta commits are relatively compact,
ranging from 163 (\cc{cc5}$\blacktriangle$)
to 1,407 changed lines (\cc{av2}$\blacktriangle$),
whereas no-CPV delta commits are substantially larger,
from 3.2K (\cc{av3}$\blacktriangle$)
to over one million lines (\cc{mg3}$\blacktriangle$).
Project build times range from 16\,s (\cc{mg3}$\blacktriangle$)
to 492\,s (\cc{ws1}$\square$),
and harness sizes from 2.4\,KB (\cc{av3}$\blacktriangle$)
to over 20\,GB (\cc{ws1}$\blacktriangle$).
See \cite{sok-website} for more details.

\PP{Challenge Project Vulnerabilities}
The 63 CPVs (33 in full-mode and 30 in delta-mode) split
into 40~C and 23~Java vulnerabilities,
covering 34 unique CWE types
including memory corruption, path traversal, and command injection.
Organizers also issued 13 SARIF broadcasts
(8 valid and 5 invalid)
to test CRS triage capability.
See \cite{sok-website} for full details.

%% file: tbl/cp_details/overview_table.tex
{\rowcolors{2}{}{gray!15}
\begin{table}[t]
\centering
\caption{Overview of open-source repositories
and challenge projects (CPs) in the final.
$\square$: full-mode; $\blacktriangle$: delta-mode.
$^*$5 unharnessed CPs are excluded:
freertos-kernel, jt808, lwip, openssl, sms4j.
$^\dag$Averaged across CPs.}
\label{t:cp-overview}
\setlength{\tabcolsep}{3pt}
\adjustbox{trim=0pt 0pt -4pt 0pt,clip,max width=\columnwidth}{
\begin{tabular}{@{}>{\cellcolor{white}}l l l cc c c r@{}}
\toprule
\textbf{Lang.} & \textbf{Project} & \textbf{Abbr.} & \multicolumn{2}{c}{\textbf{\# CPs}} & \textbf{\# CPVs} & \textbf{\# Harn.} & \textbf{SLOC}\rlap{$^\dag$} \\
\midrule
 & \href{https://github.com/curl/curl}{curl} & \cc{cu} &  & 5\,{\smaller$\blacktriangle$} & 6 & 17 & 240K \\
 & \href{https://code.videolan.org/videolan/dav1d}{dav1d} & \cc{da} & 1\,{\smaller$\square$} &  & 1 & 1 & 261K \\
 & \href{https://github.com/FreeRDP/FreeRDP}{freerdp} & \cc{fp} &  & 3\,{\smaller$\blacktriangle$} & 2 & 7 & 457K \\
 & \href{https://github.com/mm2/Little-CMS}{little-cms} & \cc{cm} & 1\,{\smaller$\square$} & 1\,{\smaller$\blacktriangle$} & 2 & 15 & 87K \\
 & \href{https://github.com/AOMediaCodec/libavif}{libavif} & \cc{av} &  & 2\,{\smaller$\blacktriangle$} & 1 & 8 & 44K \\
 & \href{https://github.com/libexif/libexif}{libexif} & \cc{ex} &  & 2\,{\smaller$\blacktriangle$} & 2 & 2 & 16K \\
 & \href{https://gitlab.gnome.org/GNOME/libxml2}{libxml2} & \cc{lx} &  & 1\,{\smaller$\blacktriangle$} & 1 & 11 & 201K \\
 & \href{https://github.com/cesanta/mongoose}{mongoose} & \cc{mg} & 1\,{\smaller$\square$} & 3\,{\smaller$\blacktriangle$} & 3 & 1 & 364K \\
 & \href{https://github.com/ntop/nDPI}{ndpi} & \cc{nd} & 1\,{\smaller$\square$} &  & 0 & 55 & 136K \\
 & \href{https://github.com/openssl/openssl}{openssl} & \cc{os} & 1\,{\smaller$\square$} &  & 0 & 30 & 909K \\
 & \href{https://github.com/shadowsocks/shadowsocks-libev}{shadowsocks-libev} & \cc{ss} & 1\,{\smaller$\square$} &  & 5 & 1 & 19K \\
 & \href{https://github.com/systemd/systemd}{systemd} & \cc{sd} & 1\,{\smaller$\square$} &  & 4 & 47 & 740K \\
 & \href{https://gitlab.com/wireshark/wireshark}{wireshark} & \cc{ws} & 1\,{\smaller$\square$} & 6\,{\smaller$\blacktriangle$} & 12 & 47 & 4901K \\
\multirow{-14}{*}{\textbf{C}} & \href{https://github.com/tukaani-project/xz}{xz} & \cc{xz} & 1\,{\smaller$\square$} &  & 1 & 4 & 41K \\
\midrule
 & \href{https://github.com/apache/commons-compress}{commons-compress} & \cc{cc} &  & 5\,{\smaller$\blacktriangle$} & 5 & 16 & 76K \\
 & \href{https://github.com/dcm4che/dcm4che}{dcm4che} & \cc{dc} & 1\,{\smaller$\square$} &  & 0 & 1 & 105K \\
 & \href{https://github.com/bioinformatics-ua/dicoogle}{dicoogle} & \cc{dg} & 1\,{\smaller$\square$} &  & 0 & 1 & 21K \\
 & \href{https://github.com/GoogleCloudPlatform/healthcare-data-harmonization}{\makecell[l]{healthcare-\\data-harmonization}} & \cc{hc} & 1\,{\smaller$\square$} &  & 0 & 1 & 53K \\
 & \href{https://github.com/apache/hertzbeat}{hertzbeat} & \cc{hb} & 1\,{\smaller$\square$} &  & 0 & 1 & 78K \\
 & \href{https://github.com/jhy/jsoup}{jsoup} & \cc{js} & 1\,{\smaller$\square$} &  & 0 & 2 & 36K \\
 & \href{https://github.com/apache/logging-log4j2}{logging-log4j2} & \cc{lj} &  & 1\,{\smaller$\blacktriangle$} & 1 & 1 & 54K \\
 & \href{https://github.com/apache/pdfbox}{pdfbox} & \cc{pb} & 1\,{\smaller$\square$} & 1\,{\smaller$\blacktriangle$} & 9 & 6 & 167K \\
 & \href{https://github.com/apache/poi}{poi} & \cc{po} & 1\,{\smaller$\square$} & 1\,{\smaller$\blacktriangle$} & 7 & 17 & 433K \\
\multirow{-10}{*}{\textbf{Java}} & \href{https://github.com/apache/tika}{tika} & \cc{tk} &  & 1\,{\smaller$\blacktriangle$} & 1 & 9 & 188K \\
\midrule
\rowcolor{white}
\textbf{Total} & \textbf{24}$^*$ & & \textbf{16}$^*$\,{\smaller$\square$} & \textbf{32}\,{\smaller$\blacktriangle$} & \textbf{63} & \textbf{301} & \\
\bottomrule
\end{tabular}
}
\end{table}
}

%% file: crs-intro.tex
\section{Cyber Reasoning Systems}
\label{s:crs:intro}

\autoref{t:crs-philosophy} summarizes
the background information behind each team's CRS.
While all teams built systems targeting the same CRS core capabilities
(\autoref{s:comp}),
their architectural approaches varied widely,
shaped by team expertise, resource constraints, and strategic priorities.
We briefly introduce each team's design philosophy,
providing context for understanding their technical choices
in subsequent sections.

\input{tbl/crs-philosophy}

\PP{Atlantis: Ensemble-First Design.}
\atlantis is built around the ensemble
philosophy~\cite{enfuzz,fu:autofz,atlantis-patchisland}:
any technique demonstrating unique contribution is worth incorporating,
and combining multiple independent approaches enhances overall robustness.
This leads to multiple independent bug-finding modules collaborating through
seed sharing,
and eight patching agents with diverse repair strategies.

\PP{Buttercup: Expertise-Driven Decomposition.}
\buttercup leverages domain expertise
to design deterministic workflows
that decompose challenges into well-defined subtasks,
with LLMs integrated only where traditional tools fall short.
Notably, \buttercup avoids high-end reasoning models,
believing that well-decomposed problems
paired with mid-tier models suffice.

\PP{RoboDuck: Agentic Design around Bug Candidates.}
\roboduck embodies an agentic-first philosophy,
building on a custom agent library
that maximizes autonomous LLM operation.
The entire system revolves around bug candidates:
from identification and filtering,
through PoV generation and patching,
to SARIF validation and bundling.

\PP{FuzzingBrain: Simple Architecture, Diverse LLM Strategies.}
\fuzzingbrain balances engineering effort against performance
by simplifying architectural design
while maximizing LLM strategy diversity.
Notably, over 90\% of its codebase is vibe-coded~\cite{vibe-coding}.
It implements 23 independent strategies,
each as a standalone Python script with minimal dependencies,
varying in scope, depth, and language-specific handling.

\PP{Artiphishell: Comprehensive Technical Coverage.}
\artiphishell achieves the most comprehensive technical coverage,
implementing diverse techniques across all four core capabilities.
To coordinate these techniques (53 components),
the team built a custom orchestration platform
that launches them on-demand
and facilitates inter-component communication.

\PP{BugBuster: Pragmatic Technology Choices.}
\bugbuster follows a pragmatic philosophy,
preferring simple and stable technology choices.
For bug finding,
the design is traditional fuzzing and program analysis centric,
with LLMs limited to auxiliary roles like seed generation.
When adopting academic techniques~\cite{42-bandfuzz,42-patchagent},
the team consistently simplifies them to be practical,
replacing sophisticated optimizations.

\PP{Lacrosse: DSPy-Based Multi-LLM Workflow.}
\lacrosse uses a Lisp-based task distributor to coordinate fuzzing~\cite{musliner-saso-2013,musliner2016fuzzbomb},
patching, and analysis in a multi-agent system~\cite{musliner:21}.
DSPy~\cite{dspy} manages diverse LLMs in parallel or as fallbacks,
with patch failures refining vulnerability analysis.

%% file: tbl/crs-philosophy.tex
{
\newcommand{\teamid}[1]{
  \raisebox{1.5ex}[0pt][0pt]{\hypertarget{team:#1}{}}\texttt{#1}
}
\begin{table}[t]
\centering
\caption{CRS Teams,
ordered top-to-bottom by final score (descending; see \autoref{t:score-breakdown}).
$^\dagger$: Full name ``All You Need IS A Fuzzing Brain''.
$^\ddagger$: Orchestration code only.
$^\S$: Uses LiteLLM~\cite{litellm} for multi-provider routing.
}
\label{t:crs-philosophy}
\smaller
\rowcolors{2}{}{gray!15}
\begin{adjustbox}{max width=\columnwidth}
\begin{tabular}{ll l l l l}
  \toprule
  \textbf{ID} & \textbf{Team} & \textbf{CRS} & \textbf{Bg} & \textbf{Lang}$^\ddagger$ & \textbf{LLM Lib} \\
  \midrule
  \teamid{AT} & Team Atlanta & \Latlantis & Mixed & Py,Rust & LangGraph~\cite{langgraph}$^\S$ \\
  \teamid{TB} & Trail of Bits & \Lbuttercup & Industry & Py & LangGraph \\
  \teamid{TI} & Theori & \Lroboduck & Industry & Py,Rust & Self-built$^\S$ \\
  \teamid{FB} & Fuzzing Brain$^\dagger$ & \Lfuzzingbrain & Academic & Py,Go & \X$^\S$ \\
  \teamid{SP} & Shellphish & \Lartiphishell & Academic & Py & Self-built$^\S$ \\
  \teamid{42} & 42-b3yond-6ug & \Lbugbuster & Academic & Py,Go & LangChain~\cite{langchain}$^\S$ \\
  \teamid{LC} & Lacrosse & \Llacrosse & Industry & Py,Lisp & DSPy \\
  \bottomrule
\end{tabular}
\end{adjustbox}
\end{table}
}

%% file: crs-tech.tex
\section{Taxonomy of CRS Techniques}
\label{s:crs}

\subsection{PoV Generation}
\label{s:crs:pov}
\input{crs-pov}

\subsection{Patch Generation}
\label{s:crs:patch}
\input{crs-patch}

\subsection{SARIF Validation}
\label{s:crs:sarif}
\input{crs-sarif}

\subsection{Bundling Strategy}
\label{s:crs:game}
\input{crs-bundle-game}

%% file: crs-pov.tex
\PP{Two Complementary Pipelines}
In \autoref{t:pov-overview},
two complementary discovery pipelines emerge as the first two row groups:
a \emph{Fuzzing Pipeline}
that extends traditional fuzzing with LLM-assisted components,
and an \emph{LLM-Based PoV Generation Pipeline}
that directly leverages LLMs
to identify vulnerabilities and generate exploit inputs.
The other two row groups capture
how teams couple them (\emph{Pipeline Cooperation})
and how PoVs are deduplicated and submitted (\emph{PoV Submission}).
Both pipelines reinforce each other:
fuzzing supplies coverage and inputs to LLM-based generation,
while LLM-generated outputs (even failures) seed fuzzers.

\input{tbl/pov-overview}

\PP{Fuzzing Pipeline}
Most teams (5/7) explored both pipelines,
while \bugbuster and \lacrosse focused on fuzzing alone.
Of the five,
\atlantis and \artiphishell explored most comprehensively;
\buttercup and \roboduck targeted fuzzing-seed techniques
(reuse, generation, sharing);
and \fuzzingbrain ran only parallel fuzzing.

\BB{Pre-competition corpus}
Five teams reused pre-collected corpora to bootstrap fuzzer coverage,
typically with two steps:
\ding{192} collecting and grouping seeds from public databases
(ClusterFuzz~\cite{clusterfuzz}, OSS-Fuzz~\cite{oss-fuzz}, GitHub, etc.)
before the competition;
\ding{193} pairing seeds with harnesses
by coverage-based ranking
or by similarity matching against harness names and LLM-inferred input formats.

\BB{Seed generation agent}
Six teams use LLMs to generate seeds in two scenarios:
early-stage bootstrap (analyzing harness code for input formats)
and troubleshooting (generating inputs for coverage
blockers~\cite{theori-branch-flipper}).
The agents typically incorporate conventional program analysis tools for better performance.
Interestingly, despite the goal being harness input generation,
all teams chose to have LLMs generate Python scripts
that produce inputs upon execution.
Additionally, \atlantis, \roboduck, and \artiphishell explored
generating input generators/mutators
and explicit grammars
(testlang and libFDP~\cite{libfdp} for \atlantis,
Python decoders for \roboduck,
and Nautilus~\cite{nautilus} grammars for \artiphishell).

\BB{Engine refinement}
Beyond existing fuzzers, teams
(\atlantis, \artiphishell, and \bugbuster; primarily from university research labs)
refined internal components
(feedback, oracle, dict, scheduler).
\artiphishell added \emph{semantic feedback}
by having LLMs generate IJON~\cite{ijon}-style annotations,
the only team to do so.
\atlantis and \artiphishell \emph{improved the Java sanitizers}
to strengthen fuzzer guidance toward valid PoCs.
For \emph{dictionary generation},
four teams produced fuzzing dictionaries
to break input-format barriers:
\atlantis via on-the-fly LLM prompts,
\artiphishell via AFL++~\cite{AFLplusplus-Woot20} \cc{dict2file} plus CodeQL~\cite{codeql},
and \bugbuster and \lacrosse via custom extraction.
Finally, \atlantis and \bugbuster customized scheduling
with \emph{directed fuzzing}:
\atlantis used a custom distance metric,
while \bugbuster used LLVM and WALA~\cite{wala} program slicing
to selectively instrument paths to sinks.

\BB{Concolic fuzzing}
\atlantis explored building hybrid fuzzers: SymCC~\cite{symcc}-based for C and a from-scratch engine for Java.

\BB{Parallel fuzzing}
All teams run multiple fuzzer instances in parallel
and synchronize corpora across them (except \fuzzingbrain).
Beyond the OSS-Fuzz defaults,
\atlantis, \artiphishell, \bugbuster, and \lacrosse
added custom C fuzzers (AFL++~\cite{AFLplusplus-Woot20}, libAFL~\cite{libafl}, or custom implementations),
with \atlantis also covering JVM via libAFL.

\PP{LLM-Based PoV Generation Pipeline}
Six teams explored this LLM-driven alternative to fuzzing.
The typical workflow involves two steps:
\ding{192} identifying and filtering bug candidates;
\ding{193} generating PoVs targeting these candidates.
\lacrosse performs only step \ding{192}
and feeds results to no-PoV patch generation (\autoref{s:crs:patch}) rather than PoV generation.
\buttercup skips step \ding{192},
letting its agent autonomously judge during PoV generation.

\BB{Bug candidate identification}
Five teams build agent systems combining LLMs,
static analysis tools (CodeQL~\cite{codeql}, Semgrep~\cite{semgrep}, Infer~\cite{infer}),
and predefined sink lists to identify and filter candidates.
Two distinctive filtering techniques stand out:
\roboduck uses LLM logprobs~\cite{openai-logprobs} as a token-efficient confidence signal,
exposing them to its classification agent as a tool;
\artiphishell and \lacrosse instead aggregate weighted votes~\cite{self-consistency}
across multiple tools and LLMs to rank candidates.

\BB{PoV Generation Agent}
Five teams construct PoV-generation agents over static and dynamic analysis tools.
Beyond source code, agents receive call paths, coverage, runtime logs,
and debugger access (GDB~\cite{gdb}/JDB~\cite{jdb}).
Four teams (\atlantis, \buttercup, \fuzzingbrain, \artiphishell)
inject CWE-specific guidance~\cite{cwe} to steer exploit construction.
Three (\atlantis, \roboduck, \artiphishell) further decompose generation:
a reach agent drives execution to the target sink,
and an exploit agent crafts the trigger.

\PP{Pipeline Cooperation}
Teams cooperate between pipelines in both directions, though more invest in one than the other.
On one side (LLM PoV Gen $\rightarrow$ Fuzz, five teams),
successful, failed, and intermediate results from LLM PoV generation
are shared with fuzzers,
expecting fuzzers to extend coverage
or mutate these near-solutions into actual PoVs.
On the other side (Fuzz $\rightarrow$ LLM PoV Gen, three teams),
fuzzers provide coverage information to guide LLM generation;
for teams with separate exploit agents,
fuzzer-found reached-but-unexploited inputs are also forwarded for exploitation.

\PP{PoV Submission}
All teams adopt straightforward strategies:
submit unique PoVs as soon as possible.
This simplicity stems from the scoring rules: correct
but duplicate submissions incur only time-decay penalties,
not accuracy penalties,
making early submission always preferable.
To minimize redundant submissions,
all teams implement deduplication
using crash stack traces, input hashing, sanitizer signatures, etc.
\roboduck and \fuzzingbrain further use LLMs
to group semantically equivalent PoVs.

%% file: tbl/pov-overview.tex
{\rowcolors{2}{}{gray!15}
\begin{table}[t]
\centering
\caption{PoV Generation Techniques Beyond OSS-Fuzz Defaults. \emptycirc: non-LLM; \fullcirc: LLM-enhanced; \V: present; blank: absent. Extended details: \autoref{t:fuzzing-techniques}.
}
\label{t:pov-overview}
\smaller
\adjustbox{trim=0pt 0pt 0pt 0pt,clip,max width=\columnwidth}{
\begin{tabular}{@{}>{\cellcolor{white}}l l ccccccc@{}}
  \toprule
  & & \hyperlink{team:AT}{\texttt{AT}} & \hyperlink{team:TB}{\texttt{TB}} & \hyperlink{team:TI}{\texttt{TI}} & \hyperlink{team:FB}{\texttt{FB}} & \hyperlink{team:SP}{\texttt{SP}} & \hyperlink{team:42}{\texttt{42}} & \hyperlink{team:LC}{\texttt{LC}} \\
  \midrule
  & Pre-Comp Corpus      & \fullcirc &  & \emptycirc &  & \fullcirc & \fullcirc & \emptycirc \\
  & Seed Gen Agent        & \fullcirc & \fullcirc & \fullcirc &  & \fullcirc & \fullcirc & \fullcirc \\
  & \tsp\tmid{} Bootstrap & \fullcirc & \fullcirc &  &  & \fullcirc & \fullcirc & \fullcirc \\
  & \tsp\tmid{} Solve cov blocker & \fullcirc & \fullcirc & \fullcirc &  & \fullcirc &  &  \\
  & \tsp\tmid{} Mutator/generator & \fullcirc &  &  &  &  &  &  \\
  & \tsp\tlast{} Grammar-aware & \fullcirc &  & \fullcirc &  & \fullcirc &  &  \\
  & Engine Refinement     & \fullcirc &  &  &  & \fullcirc & \emptycirc & \emptycirc \\
  & \tsp\tmid{} Semantic feedback &  &  &  &  & \fullcirc &  &  \\
  & \tsp\tmid{} Improved sanitizer & \emptycirc &  &  &  & \emptycirc &  &  \\
  & \tsp\tmid{} Dict Gen & \fullcirc &  &  &  & \emptycirc & \emptycirc & \emptycirc \\
  & \tsp\tlast{} Directed fuzzing & \emptycirc &  &  &  &  & \emptycirc &  \\
  & Concolic Fuzzing     & \emptycirc &  &  &  &  &  &  \\
  & Parallel Fuzzing           & \emptycirc & \emptycirc & \emptycirc & \emptycirc & \emptycirc & \emptycirc & \emptycirc \\
  & \tsp\tmid{} Corpus sync & \emptycirc & \emptycirc & \emptycirc &  & \emptycirc & \emptycirc & \emptycirc \\
  & \tsp\tmid{} Added C fuzzers & \emptycirc &  &  &  & \emptycirc & \emptycirc & \emptycirc \\
  \multirow{-16}{*}{\parbox{0.8cm}{\raggedright\textbf{Fuzzing Pipeline}}}
  & \tsp\tlast{} Added JVM fuzzers & \emptycirc &  &  &  &  &  &  \\
  \midrule
  & Bug Cand.\ I.D.      & \fullcirc &  & \fullcirc & \fullcirc & \fullcirc &  & \fullcirc \\
  & \tsp\tmid{} Candidate filter & \fullcirc &  & \fullcirc & \fullcirc & \emptycirc &  & \fullcirc \\
  & \tsp\tlast{} Non-PoV Gen usage  & \V &  & \V &  & \V &  & \V \\
  & PoV Gen Agent             & \fullcirc & \fullcirc & \fullcirc & \fullcirc & \fullcirc &  &  \\
  & \tsp\tmid{} With CWE guidance    & \fullcirc & \fullcirc &  & \fullcirc & \fullcirc &  &  \\
  \multirow{-6}{*}{\parbox{0.8cm}{\raggedright\textbf{LLM-Based PoV Gen Pipeline}}}
  & \tsp\tlast{} Reach-then-exploit & \fullcirc &  & \fullcirc &  & \fullcirc &  &  \\
  \midrule
  & LLM PoV Gen $\rightarrow$ Fuzz & \V & \V & \V & \V & \V &  &  \\
  \multirow{-2}{*}{\parbox{0.8cm}{\raggedright\textbf{Pipeline Co-op}}}
  & Fuzz $\rightarrow$ LLM PoV Gen & \V &  & \V &  & \V &  &  \\
  \midrule
  & Deduplication        & \emptycirc & \emptycirc & \fullcirc & \fullcirc & \emptycirc & \emptycirc & \emptycirc \\
  \multirow{-2}{*}{\parbox{0.8cm}{\raggedright\textbf{PoV Sub.}}}
  & ASAP Submission        & \V & \V & \V & \V & \V & \V & \V \\
  \bottomrule
\end{tabular}
}
\end{table}
}

%% file: crs-patch.tex
All CRSs follow a de facto patch pipeline as shown below,
where RCA denotes Root Cause Analysis
and brackets indicate optional steps.

\begin{center}
\small\texttt{loop([RCA] → Generate → Validate) → Dedup → Submit}
\end{center}

Within this pipeline,
teams explore different LLM-centric designs to make patching effective.
\autoref{t:patch-overview} organizes their design choices
into five row groups.
\emph{Agent Arch.} captures the overall agentic design
of the patch system.
The other four mirror the pipeline steps: \emph{RCA},
\emph{Generation}, \emph{Validation}, and \emph{Dedup. \& Sub.}
The last two steps are grouped together
because deduplication primarily prepares for submission.

\input{tbl/patch-overview}

\PP{Agent Architecture}
CRSs' designs split into three patterns.

\BB{Multi-Arch}
Multi-Arch is an ensemble strategy where the CRS patch system runs multiple distinct patcher architectures to combine their benefits.
\atlantis ensembles eight patcher agents~\cite{atlantis-patchisland} spanning diverse designs:
workflow-based pipelines with ReAct-style~\cite{yao2023react} tool use,
autonomous agents with iterative context retrieval,
multi-agent systems for handling context limitations,
and off-the-shelf coding agents (Aider~\cite{aider}, SWE-Agent~\cite{sweagent}).
\artiphishell instead pairs a fully agentic LLM patcher with a program-analysis-assisted, one-shot minimal LLM patcher.
When generating,
\atlantis stops once any agent produces a valid patch,
while \artiphishell retains all candidates,
then ranks and strategically selects the final patches
(see \emph{Deduplication and Submission} below).

\BB{Multi-Agent}
Multi-Agent is a decomposition strategy where a single patcher contains multiple coordinating sub-agents.
\buttercup organizes its four agents into a pipeline of RCA, fix strategy, patch creation, and reflection, each handing off to the next.
\roboduck instead uses a hierarchical design:
a ReAct-style outer loop generates patches,
invoking SourceQuestionsAgent as an inner tool for code understanding.

\BB{Single-Agent}
Three CRSs adopt single-agent designs
with varying degrees of agentic customization.
\fuzzingbrain implements 23 strategies
spanning delta-scan and full-scan modes,
differentiated by context scope and knowledge injection.
\bugbuster maximizes configuration diversity,
exploring 16 combinations of temperature and prompt context (failed cases, stack
traces, etc) within a single agent architecture.
\lacrosse uses a DSPy-based workflow
with model escalation from cheaper to expensive models.

\PP{Root Cause Analysis}
Four CRSs (\atlantis, \buttercup, \roboduck, \artiphishell)
implement standalone RCA components,
allowing LLMs to separately focus on
root cause analysis and patch synthesis as distinct sub-problems.
All four build LLM agents for RCA;
\buttercup, \roboduck, and \artiphishell
further leverage information from multiple PoVs
for more accurate root cause analysis.
\artiphishell additionally incorporates a non-LLM RCA component
that combines multi-source signals
(SAST reports, stack traces, fuzzing invariants, etc.)
with weighted voting to rank root cause candidates.

\PP{Generation}
Within the Generate step, CRSs employ various techniques
to improve patch quality.

\BB{Contextualization}
This presents the information sources supplied to the LLM's prompt for patch generation.
The crash/sanitizer output and agentic code search over project code are used by all teams.
Optional augmentations, by descending popularity:
pre-built code indexers for symbol lookups (5 teams);
SAST reports static analysis outcomes (4 teams);
runtime probes (debugger, coverage instrumentation) for inspecting actual execution (4 teams);
the PoV bytes that triggered the bug (3 teams);
CWE-specific vulnerability domain knowledges (2 teams);
and LLM fine-tuning (Llama~\cite{llama}) for context retrieval (1 team).

\BB{LLM Reflection}
LLM reflection~\cite{agent-reflexion} enables agents to learn from failed attempts;
five CRSs adopted it.
One typical example is \buttercup,
which implements a dedicated reflection agent
that analyzes failures at each generation step
and provides corrective guidance.

\BB{No-PoV Patch Generation}
Three CRSs (\roboduck, \fuzzingbrain, \lacrosse)
attempt patch generation for a vulnerability candidate
whose PoV has not been created by the CRS.
Without dynamic evidence, this approach carries risk,
but can address vulnerabilities that are obvious to identify
yet difficult to trigger with a PoV.
The generation technique is similar to but more limited than PoV-based
techniques,
while all teams focus on risk mitigation
by limiting no-PoV patch quantities per challenge
and imposing stricter submission conditions
(\eg, delayed submission, gated by prior success rate, etc).

\PP{Validation}
CRSs employ several checks within each iteration
before accepting a candidate patch.

\BB{Basic Checks}
All CRSs share three basic checks:
build verification,
PoV reproduction during generation (\emph{PoV Test (Gen)}),
and project test suites
(\bugbuster skips project tests).
The number of PoVs used during generation varies (1/N/* in \autoref{t:patch-overview});
most teams use only a subset to keep the iteration loop fast,
then revalidate against more before submission
(see \emph{PoV Test (Submit)} below).

\BB{PoV Test (Submit)}
Before submission, most CRSs revalidate patches
against a broader PoV set than was used during generation,
catching incomplete fixes that a partial generation-time set may miss.
\atlantis and \fuzzingbrain expand from a single PoV in generation
to multiple PoVs before submission,
\buttercup uses multiple throughout,
and \lacrosse uses a single PoV throughout.

\BB{LLM-as-Judge}
Three CRSs incorporate LLM-based evaluation~\cite{llm-judge},
including judging whether patches correctly address the root cause
and follow the prescribed fix strategy (\atlantis),
and self-reflecting on whether patches genuinely fix vulnerabilities
rather than being superficial, easily-bypassed, or having side effects
(\fuzzingbrain/\artiphishell for No-PoV/all patches).

\BB{Post-patch Fuzz}
\fuzzingbrain and \artiphishell adopt short-term fuzzing
on patched projects for incomplete patch detection.

\BB{Rebuild Optimization}
\atlantis and \artiphishell employ build caching
to accelerate iterative patch refinement
(ccache~\cite{ccache} for C/C++, Maven~\cite{maven} caching for Java).

\PP{Deduplication and Submission}
Naively submitting a patch for each PoV
incurs many duplicate submissions and patch-score penalties,
so teams must balance deduplication and submission timing.
Since the two strategies are tightly coupled,
we discuss them together.

\BB{Minimal Patch Set Calculation}
Four CRSs (\atlantis, \buttercup, \artiphishell, and \bugbuster)
leverage the fact that a single patch can fix multiple PoVs
sharing the same root cause,
and compute a minimal patch set that covers all known PoVs
to avoid duplicate submissions.
These CRSs differ in three dimensions:
\ding{192}~\emph{Calc. timing}:
on each new PoV (\atlantis, \buttercup, \artiphishell) for faster response,
or hourly (\bugbuster) for better global optimization.
\ding{193}~\emph{Calc. mode}:
incremental over uncovered PoVs only (\atlantis, \bugbuster) for simplicity,
or recompute over all PoVs (\buttercup, \artiphishell) for better optimization
at the risk of duplicate submissions.
\ding{194}~\emph{Submission timing}:
immediate (\atlantis, \buttercup, \bugbuster),
or delayed (\artiphishell, $\geq$60min) for better global minima.
For \artiphishell, the candidates retained from its Multi-Arch ensemble
(\autoref{s:crs:patch}) are submitted ranked by PoV count.

\BB{No-PoV Patch Delayed Submission}
All three CRSs with No-PoV patch generation capability
(\roboduck, \fuzzingbrain, \lacrosse)
delay submission to reduce imperfect patch penalties:
\roboduck waits $\geq$45min and gates on PoV patch success history,
\fuzzingbrain waits until 50\% of challenge time,
and \lacrosse submits 30min before deadline.

%% file: tbl/patch-overview.tex
{\rowcolors{2}{}{gray!15}
\begin{table}[t]
\centering
\caption{Patch Generation Techniques. \fullcirc: present; blank: no custom implementation; --: not applicable; 1/N/$*$: single/multiple/all PoVs; $^\dagger$all teams use sanitizer/crash reports and failed patch feedback. Extended details: \autoref{t:crs-patch-techniques}.}
\label{t:patch-overview}
\smaller
\adjustbox{trim=0pt 0pt 0pt 0pt,clip,max width=\columnwidth}{
\begin{tabular}{@{}>{\cellcolor{white}}l l lllllll@{}}
  \toprule
  & & \hyperlink{team:AT}{\texttt{AT}} & \hyperlink{team:TB}{\texttt{TB}} & \hyperlink{team:TI}{\texttt{TI}} & \hyperlink{team:FB}{\texttt{FB}} & \hyperlink{team:SP}{\texttt{SP}} & \hyperlink{team:42}{\texttt{42}} & \hyperlink{team:LC}{\texttt{LC}} \\
  \midrule
  \rowcolor{gray!15} & Multi-Arch & \fullcirc &  &  &  & \fullcirc &  &  \\
  \rowcolor{gray!15} & Multi-Agent &  & \fullcirc & \fullcirc &  &  &  &  \\
  \rowcolor{gray!15} \multirow{-3}{*}{\parbox{0.9cm}{\raggedright\textbf{Agent Arch.}}}
  & Single-Agent &  &  &  & \fullcirc &  & \fullcirc & \fullcirc \\
  \midrule
  & Standalone RCA & \fullcirc & \fullcirc & \fullcirc &  & \fullcirc &  &  \\
  & \tsp\tmid{} Multi-PoV RCA &  & \fullcirc & \fullcirc &  & \fullcirc &  &  \\
  \multirow{-3}{*}{\parbox{0.8cm}{\raggedright\textbf{RCA}}}
  & \tsp\tlast{} Non-LLM RCA &  &  &  &  & \fullcirc &  &  \\
  \midrule
  & Contextualization$^\dagger$ & \fullcirc & \fullcirc & \fullcirc & \fullcirc & \fullcirc & \fullcirc & \fullcirc \\
  & \tsp\tmid{} Code Indexer & \fullcirc & \fullcirc & \fullcirc &  & \fullcirc & \fullcirc &  \\
  & \tsp\tmid{} SAST & \fullcirc &  & \fullcirc & \fullcirc & \fullcirc &  &  \\
  & \tsp\tmid{} CWE Guidance &  &  &  & \fullcirc &  & \fullcirc &  \\
  & \tsp\tmid{} Fine-tuned LLM & \fullcirc &  &  &  &  &  &  \\
  & \tsp\tmid{} Agentic Code Search & \fullcirc & \fullcirc & \fullcirc & \fullcirc & \fullcirc & \fullcirc & \fullcirc  \\
  & \tsp\tmid{} Dynamic Info & \fullcirc &  & \fullcirc & \fullcirc & \fullcirc &  &  \\
  & \tsp\tlast{} PoV Bytes & \fullcirc &  & \fullcirc &  &  &  & \fullcirc \\
  & LLM Reflection & \fullcirc & \fullcirc & \fullcirc &  & \fullcirc &  & \fullcirc \\
    \multirow{-10}{*}{\parbox{0.8cm}{\raggedright\textbf{Gen\-er\-ation}}}
  & No-PoV Patch Generation &  &  & \fullcirc & \fullcirc &  &  & \fullcirc \\
  \midrule
  & Basic Checks & \fullcirc & \fullcirc & \fullcirc & \fullcirc & \fullcirc & \fullcirc & \fullcirc \\
  & \tsp\tmid{} Build & \fullcirc & \fullcirc & \fullcirc & \fullcirc & \fullcirc & \fullcirc & \fullcirc \\
  & \tsp\tmid{} PoV Test (Gen) & 1 & N & $*$ & 1 & N & $*$ & 1 \\
  & \tsp\tlast{} Proj. Tests & \fullcirc & \fullcirc & \fullcirc & \fullcirc & \fullcirc &  & \fullcirc \\
  & PoV Test (Submit) & $*$ & N & $*$ & N & $*$ & $*$ & 1 \\
  & LLM as Judge & \fullcirc &  &  & \fullcirc & \fullcirc &  &  \\
  & Post-patch Fuzz &  &  &  & \fullcirc & \fullcirc &  &  \\
  \multirow{-8}{*}{\parbox{0.8cm}{\raggedright\textbf{Val\-i\-da\-tion}}}
  & Rebuild Optimization & \fullcirc &  &  &  & \fullcirc &  &  \\
  \midrule
  & Min. Patch Set Calc. & \fullcirc & \fullcirc &  &  & \fullcirc & \fullcirc &  \\
  \multirow{-2}{*}{\parbox{0.8cm}{\raggedright\textbf{Dedup. \& Sub.}}}
  & No-PoV Delayed Sub. & -- & -- & \fullcirc & \fullcirc & -- & -- & \fullcirc \\
  \bottomrule
\end{tabular}
}
\end{table}
}

%% file: crs-sarif.tex
\input{tbl/sarif-submission-strategy}

The SARIF validation task requires CRSs to assess
each static analysis report as valid or invalid,
submitting a verdict of \emph{Correct} or \emph{Incorrect}.
CRSs can resubmit to revise verdicts while incurring penalties in both time and
accuracy.

\PP{Key Evidence for Validation}
\autoref{t:sarif-submission} summarizes each team's core submission strategy.
Teams primarily relied on three types of evidence:
\ding{192}~\cc{Match Any PoV}: matching SARIF locations against crash
information from exploited vulnerabilities;
\ding{193}~\cc{Match Bug Cand.}: matching any bug candidate (inferred from LLM,
static analysis, PoV, etc.);
\ding{194}~\cc{LLM As Judge}: agentic prompting to directly assess its
correctness.

\PP{Validation Strategies}
Teams adopted three strategies.

\BB{PoV-centric}
\atlantis, \buttercup, and \fuzzingbrain primarily rely on PoV matching,
submitting \emph{Correct} only when a match is found
and withholding unmatched reports.
\fuzzingbrain additionally uses a fallback LLM judgement,
but only submits \emph{Correct} from it.

\BB{LLM-judge-centric}
\artiphishell, \bugbuster, and \lacrosse rely on LLM judgement,
submitting both \emph{Correct} and \emph{Incorrect} based on the model's
assessment.
\bugbuster falls back to PoV matching when the model replies with uncertainty.

\BB{Bug-cand-centric}
\roboduck matches SARIF reports against its bug candidate database,
initially submitting \emph{Incorrect} for unmatched reports
and revising to \emph{Correct} on new evidence.

%% file: tbl/sarif-submission-strategy.tex
\begin{table}[t]
\centering
\small
\caption{SARIF Submission Strategies. \V/\X: submit \emph{Correct}/\emph{Incorrect}. Full details at \cite{sok-website}.
}
\label{t:sarif-submission}
\adjustbox{trim=0pt 0pt 0pt 0pt,clip,max width=\columnwidth}{
\rowcolors{2}{}{gray!15}
\begin{tabular}{@{}>{\centering\arraybackslash}m{1.2cm}@{\hspace{0.6em}}m{2.8cm}@{\hspace{0.6em}}l@{}}
\toprule
\textbf{CRS} & \textbf{Category} & \multicolumn{1}{l}{\textbf{Submission Strategy Overview}} \\
\midrule
\hyperlink{team:AT}{\texttt{AT}}, \hyperlink{team:TB}{\texttt{TB}} & PoV-centric &
\begin{tikzpicture}[baseline=(current bounding box.center), node distance=0.5cm, >=stealth, font=\small]
  \node[text width=2cm, align=center] (pov) {\cc{Match Any PoV}};
  \node[text width=0.8cm, align=center] (submit) [right=1.2cm of pov] {\cc{\V}};
  \draw[->] (pov) -- node[above, inner sep=2pt] {\emph{Y}} (submit);
  \draw[->] ([xshift=-0.3cm]pov.north) -- ++(0,0.25) -| node[pos=0.25, above, inner sep=2pt] {\emph{N}} ([xshift=-0.15cm]pov.west) -- (pov.west);
\end{tikzpicture} \\[1.5ex]
\midrule
\hyperlink{team:FB}{\texttt{FB}} & PoV-centric &
\begin{tikzpicture}[baseline=(current bounding box.center), node distance=0.5cm, >=stealth, font=\small]
  \node[text width=2cm, align=center] (pov) {\cc{Match Any PoV}};
  \node[text width=0.8cm, align=center] (submit1) [right=1.2cm of pov] {\cc{\V}};
  \node[text width=2cm, align=center, below=0.5cm of pov] (llm) {\cc{LLM As Judge}};
  \node[text width=0.8cm, align=center] (submit2) [right=1.2cm of llm] {\cc{\V}};
  \draw[->] (pov) -- node[above, inner sep=2pt] {\emph{Y}} (submit1);
  \draw[->] (pov) -- node[right, inner sep=2pt] {\emph{N}} (llm);
  \draw[->] (llm) -- node[above, inner sep=2pt] {\emph{Y}} (submit2);
  \draw[->] ([xshift=0.2cm]llm.west) -- ++(-0.5,0) |- node[pos=0.25, right, inner sep=2pt] {\emph{N}} ([xshift=0.2cm]pov.west);
\end{tikzpicture} \\[2.5ex]
\midrule
\hyperlink{team:TI}{\texttt{TI}} & Bug-cand-centric &
\begin{tikzpicture}[baseline=(current bounding box.center), node distance=0.5cm, >=stealth, font=\small]
  \node[text width=2cm, align=center] (cand1) {\cc{Match Bug Cand.}};
  \node[text width=1cm, align=center] (submit1) [right=1.2cm of cand1] {\cc{\V/\X}\\{(1st)}};
  \node[text width=2cm, align=center, below=0.5cm of cand1] (cand2) {\cc{Match Bug Cand.}};
  \node[text width=1cm, align=center] (resub) [right=1.2cm of cand2] {\cc{\V}\\{(2nd)}};
  \draw[->] (cand1) -- node[above, inner sep=2pt] {\emph{Y/N}} (submit1);
  \draw[->] (cand1) -- node[right, inner sep=2pt] {\emph{N}} (cand2);
  \draw[->] (cand2) -- node[above, inner sep=2pt] {\emph{Y}} (resub);
  \draw[->] ([xshift=-0.3cm]cand2.north) -- ++(0,0.2) -| node[pos=0.25, above, inner sep=2pt] {\emph{N}} ([xshift=-0.15cm]cand2.west) -- (cand2.west);
\end{tikzpicture} \\[2.5ex]
\midrule
\hyperlink{team:SP}{\texttt{SP}}, \hyperlink{team:LC}{\texttt{LC}} & LLM-judge-centric &
\begin{tikzpicture}[baseline=(llm.center), node distance=0.5cm, >=stealth, font=\small]
  \node[text width=2cm, align=center] (llm) {\cc{LLM As Judge}};
  \node[text width=0.8cm, align=center] (submit) [right=1.2cm of llm] {\cc{\V/\X}};
  \draw[->] (llm) -- node[above, inner sep=2pt] {\emph{Y/N}} (submit);
\end{tikzpicture} \\[1.5ex]
\midrule
\hyperlink{team:42}{\texttt{42}} & LLM-judge-centric &
\begin{tikzpicture}[baseline=(current bounding box.center), node distance=0.5cm, >=stealth, font=\small]
  \node[text width=2cm, align=center] (llm) {\cc{LLM As Judge}};
  \node[text width=0.8cm, align=center] (submit) [right=1.2cm of llm] {\cc{\V/\X}};
  \node[text width=2cm, align=center, below=0.5cm of llm] (pov) {\cc{Match Any PoV}};
  \node[text width=0.8cm, align=center] (submit2) [right=1.2cm of pov] {\cc{\V}};
  \draw[->] (llm) -- node[above, inner sep=2pt] {\emph{Y/N}} (submit);
  \draw[->] (llm) -- node[right, inner sep=2pt] {\emph{UNK}} (pov);
  \draw[->] (pov) -- node[above, inner sep=2pt] {\emph{Y}} (submit2);
  \draw[->] ([xshift=-0.3cm]pov.north) -- ++(0,0.2) -| node[pos=0.25, above, inner sep=2pt] {\emph{N}} ([xshift=-0.15cm]pov.west) -- (pov.west);
\end{tikzpicture} \\
\bottomrule
\end{tabular}
}
\end{table}

%% file: crs-bundle-game.tex
Bundling pairs PoVs, patches, and SARIF assessments
into coherent vulnerability reports.
Unlike other submissions with time decay,
bundling allows free updates until the deadline,
with scoring based solely on final results.
A bundle can contain any two of three pairings, PoV-Patch, PoV-SARIF, and Patch-SARIF, to
form a complete scoring bundle,
while any incorrect pairing will penalize the entire bundle.
\input{tbl/bundling-overview}

\autoref{t:bundling-overview} summarizes each team's bundling pairing strategies.
Given the risk of score penalties,
teams tend to derive pairings from existing workflows
rather than inferring relationships independently.
\ding{192}
All teams naturally derive PoV-Patch relationships from PoV-based patch generation.
For No-PoV patches~(\autoref{s:crs:patch}),
\roboduck retroactively links PoVs once discovered,
while \fuzzingbrain does not.
\ding{193}
For SARIF pairings,
teams either reuse their SARIF validation results~(\autoref{s:crs:sarif})
or use SARIF reports to generate PoVs/patches,
pairing them upon success.
\bugbuster and \lacrosse do not participate in SARIF pairing,
while only teams with No-PoV patch capability can submit Patch-SARIF bundles.

%% file: tbl/bundling-overview.tex
{\rowcolors{2}{}{gray!15}
\begin{table}[t]
\centering
\caption{Bundling Pairing Strategies. \fullcirc: used; blank: not used.}
\label{t:bundling-overview}
\smaller
\adjustbox{trim=0pt 0pt 0pt 0pt,clip,max width=\columnwidth}{
\begin{tabular}{@{}>{\cellcolor{white}}l l lllllll@{}}
  \toprule
  \textbf{Pairing} & \textbf{Source} & \hyperlink{team:AT}{\texttt{AT}} & \hyperlink{team:TB}{\texttt{TB}} & \hyperlink{team:TI}{\texttt{TI}} & \hyperlink{team:FB}{\texttt{FB}} & \hyperlink{team:SP}{\texttt{SP}} & \hyperlink{team:42}{\texttt{42}} & \hyperlink{team:LC}{\texttt{LC}} \\
  \midrule
  & PoV-based Patch & \fullcirc & \fullcirc & \fullcirc & \fullcirc & \fullcirc & \fullcirc & \fullcirc \\
  \multirow{-2}{*}{PoV-Patch} & Match No-PoV Patch &  &  & \fullcirc &  &  &  &  \\
  \midrule
  & SARIF Validation & \fullcirc & \fullcirc & \fullcirc & \fullcirc &  &  &  \\
  \multirow{-2}{*}{PoV-SARIF} & SARIF-guided PoV &  &  &  & \fullcirc & \fullcirc &  &  \\
  \midrule
  & Bug Candidate DB &  &  & \fullcirc &  &  &  &  \\
  \multirow{-2}{*}{Patch-SARIF} & SARIF-guided Patch &  &  &  & \fullcirc &  &  &  \\
  \bottomrule
\end{tabular}
}
\end{table}
}

%% file: ana.tex
\section{Competition Result Analysis}
\label{s:analysis}

\input{tbl/competition-results}

\subsection{What Scores Reveal (and Conceal)}
\label{s:analysis:score}

We analyze the final scores
(\autoref{f:score-over-time}, \autoref{t:score-breakdown})
to understand what differentiated the finalists.
The competition spanned 142.7 hours across seven phases (P1--P7),
with tasks released concurrently within each phase.\footnote{Phases are non-overlapping time windows during which CPs are released to teams; the CP-to-phase assignment has no intentional structure.}

\PP{System stability.}
Stability accounts for most of the score gap.
As \autoref{f:score-over-time} shows,
three teams experienced severe stability issues at this scale of autonomous evaluation.
\atlantis (392.8 points, nearly 80\% more than second-place \buttercup at 219.4)
sustained activity across all phases, while
\buttercup and \roboduck were competitive early
but plateaued after P3 and P4 respectively,
visibly dropping out of contention.
\lacrosse, unfortunately, contributed only in early phases
and was largely inactive afterward.

Pinpointing the cause of each stability failure is inherently difficult:
CRSs are highly parallel distributed systems,
and teams selectively uploaded telemetry to the organizers.
The most likely hypotheses are as follows.
\buttercup and \roboduck's plateaus likely share a common trigger
in P3's \cc{wireshark} CP,
which required about 1\,TB of disk for compiled artifacts
(20.9\,GB across 47 harnesses)
and can crash CRSs lacking robust disk management.
We also found a CP-cleanup issue in \roboduck's telemetry
(reporting P1 CP LLM activities during P3),
which could have worsened the situation.
\lacrosse's postmortem identifies
an unrecoverable OOM crash on their master scheduler node,
after which their system stopped working
and the only output was 1,200+ PoVs against a single CPV.

Even teams that reported activity throughout were not free from stability bugs.
Per their own analyses,
\bugbuster's submission bug~\cite{42-open-letter}
cost most of its patch points (14.2),
\artiphishell reported multiple system issues in their postmortem~\cite{shellphish-pm},
such as incorrect LLM budget configuration,
and \atlantis silently failed on all \cc{poi} Java challenges (heartbeat telemetry activities only).

\PP{Submission accuracy.}
Accuracy was another major factor in the final ranking
(\autoref{t:game-strategy}):
the competition's accuracy multiplier (\autoref{s:comp})
directly scales each team's total by their per-challenge accuracy rate.
While most CRSs reach high submission accuracy
(PoV $>$80\%, Patch $>$75\%),
\roboduck and \fuzzingbrain show significantly lower rates
(PoV at 61.8\% and 80.0\%, Patch at 31.7\% and 23.3\% respectively),
mostly driven by bursts of invalid submissions on a few challenges
rather than a uniformly weak pipeline
(\autoref{s:analysis:patch}).
These result in the two largest accuracy penalties ($-16.3$ and $-13.5$),
and for \roboduck the penalty was decisive:
its pre-penalty score was higher than \buttercup's,
but the penalty dropped \roboduck to third,
behind \buttercup by 8.7 points.

\PP{Technical capability.}
A direct comparison of technical capability would be the most interesting result here,
but the factors discussed above undermine the score-based conclusion validity.
A low score may indicate a real capability gap,
or it may simply reflect outages or strategy choices,
and we cannot tell them apart.
A high score is informative to some extent:
the team's design must have had corresponding effectiveness.
Reading the leaders on this basis (\autoref{t:score-breakdown}),
\atlantis shows strong performance on C PoV (52.6),
Patch (171.0), and Bundle (136.2);
\roboduck on Java PoV (31.6);
\bugbuster on SARIF (9.7);
and \artiphishell on per-submission accuracy (smallest penalty, $-0.1$).

\input{tbl/score-breakdown}

\PP{The AIxCC trifecta}
AIxCC is fundamentally a test of three intertwined capabilities:
\emph{research}, \emph{engineering}, and \emph{strategy}.
Winning required balancing all three,
and the finalists illustrate contrasting trade-offs.
\roboduck built the most agentic architecture
with advanced technique designs,
but its aggressive strategy led to the largest accuracy penalty.
\fuzzingbrain, among the smallest teams,
combined a focused strategy with vibe coding and LLM-assisted expertise
to ship an effective system on minimal resources.
\artiphishell and \atlantis pursued similar broad-coverage approaches
(both large research teams investing heavily in engineering
and exploring many research directions),
with different execution outcomes.
\artiphishell encountered reliability issues
that limited how much of its broad coverage reached the scoreboard,
while \atlantis's ensemble executed reliably,
sustaining scoring across all seven phases.
In all cases,
technique capability alone did not decide the outcome;
strategy and engineering shaped it as much.

\findingbox{Key Finding (KF) 1.}{Winning AIxCC requires balancing research, engineering, and strategy. Stability proved the most fundamental requirement, yet many teams failed.}

\begin{figure*}[t!]
    \centering
    \includegraphics[width=\textwidth]{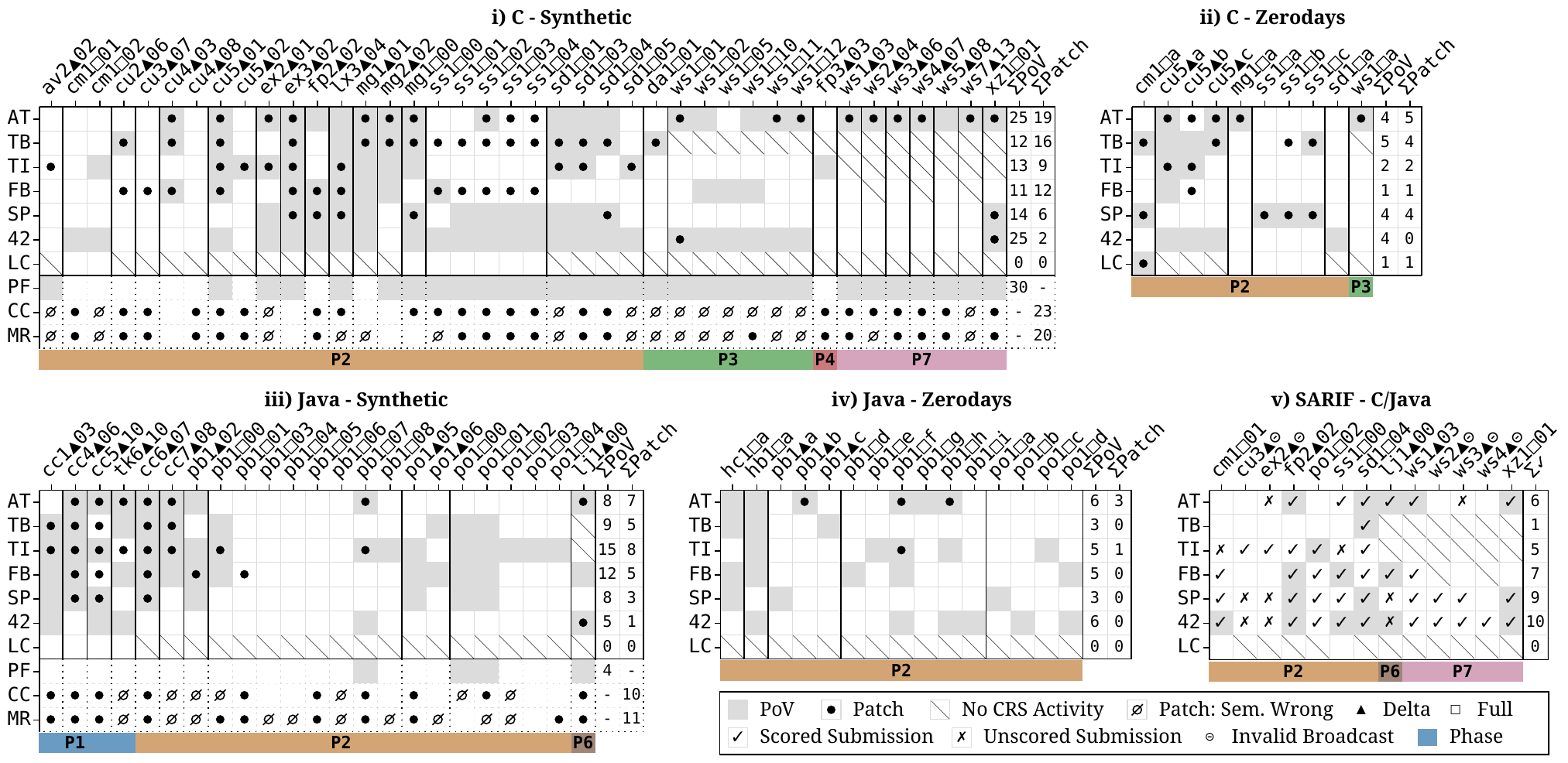}
    \caption{Team performance per CPV.
  Matrices i)--iv): detected/patched CPVs and 0-days; v): SARIF assessment.
  Diagonal: no CRS log messages.
  PF/CC/MR rows are annotations of 3-run union results.
  In patch: \fullcirc~valid; \ensuremath{\varnothing}/blank: failed manual/automated validation.
  SARIF: invalid (\scalebox{0.7}{\textcircled{-}}) expects \emph{Incorrect}; {\scriptsize \ding{51}}/{\scriptsize \ding{55}}~= CRS judgment \emph{Incorrect}/\emph{Correct};
  CPV abbreviations in \autoref{s:app:abbr}.
  }
    \label{fig:team-performance-per-cpv}
\end{figure*}

\subsection{Auxiliary CPV Annotation}
\label{s:analysis:annotation}

\PP{Unknown CRS Capability Boundaries}
Beyond surface-level score comparison,
we want deeper insight into CRS performance on each CPV.
However, per-CPV analysis faces a key question:
where do CRSs' true capability boundaries lie?
This is hard to answer because a CRS failure on a CPV
can stem from either a capability limit or an engineering/strategy issue,
and the existing logs/telemetry infrastructure
cannot support such accurate root-cause analysis
(as detailed in \autoref{s:analysis:score}).
We therefore shift the angle:
rather than diagnosing CRSs directly,
we characterize the CPVs through independent experiments.
Specifically, we run representative bug-finding and patching techniques
on each CPV under ideal laboratory conditions with competition-level
resources, annotating what should be CRS-solvable.
Thus, comparing CRS outcomes against this reference exposes
advances beyond representative techniques,
obstacles on CPVs that should have been solvable,
and boundaries that current techniques cannot cross.

\PP{Annotation Techniques}
To annotate which CPVs CRSs should be able to solve,
we select techniques every finalist CRS has invested in, per \autoref{s:crs}.
For bug finding, we use parallel fuzzing (PF),
the de facto standard non-LLM bug-finding approach.
For patching, we use MultiRetrieval (MR)~\cite{kim:atlantis},
a minimal agentic code-search-and-patch loop.
We also evaluate Claude Code (CC)~\cite{claude-code}, the most capable
general-purpose coding agent, as a patching reference.

This annotation setup is an approximation of what CRSs should solve,
not an exhaustive characterization of their technical boundary.
Specifically:
\ding{192}~each technique runs in an ideal lab environment,
isolated from end-to-end system-level challenges
(scheduling, multi-harness coordination, PoV/patch deduplication/pairing);
\ding{193}~resource budgets approximate a reasonable CRS resource allocation
(per-harness bug-finding budget; per-crash patching budget);
\ding{194}~the LLM used in annotation is contamination-free and accessible during competition
(\cc{claude-3-7-sonnet-20250219}, training cutoff October 2024, predating
AIxCC's final challenge); CC uses the closest still-available release.

\emph{PF (\cc{atlantis-multilang-given_fuzzer}~\cite{oss-crs}).}
PF runs on vulnerable harnesses using 16 cores each,
with shared-memory seed sharing and OSS-Fuzz configurations from the final competition.
Each run lasts up to 6 hours per harness,
stopping early when all organizer-synthesized CPVs in that harness are found.
Total: 8,906 CPU-hours across 3 runs.

\emph{MR (from \atlantis, in OSS-CRS~\cite{oss-crs}).}
It is a minimal patch agent using AST-based code search~\cite{tree-sitter}.
It receives ground-truth PoVs, sanitizer reports, and crash stacks,
generating up to 3 patches per run across 3 runs per CPV.
Patches are validated automatically (build, PoV reproduction, functional tests),
then cross-validated by two experts for semantic correctness.
Cost: \$221.69 (LLM), one person-week (manual cross-validation).

\emph{CC (v1.0.88, 2025-08-21).}
A general-purpose coding agent; otherwise per MR.
Cost: \$119.53 (LLM), one person-week.

\PP{Annotated Overview}
\autoref{fig:team-performance-per-cpv} presents per-CPV outcomes for all CRSs
alongside our annotations.
Overall, the annotation results reflect that
the final's challenges do not primarily reward
solving exceptionally difficult problems:
around half are solvable by a single annotation technique
(PF: 34/63 PoV; MR: 31/63 patches; CC: 33/63 patches).

Combining these annotations with CRS performance,
two contrasting patterns emerge.
On one hand, most CRSs demonstrate genuine improvements
beyond annotation techniques,
finding and patching many non-annotated CPVs,
particularly in P1--P2 C and Java challenges
where LLM-based reasoning proves essential.
On the other hand, CRSs also unexpectedly underperform
on annotated solvable challenges.
The causes are multifaceted:
system-wide failures (\buttercup, \roboduck ceased after P3--P4),
critical bugs (\bugbuster's patch submission issue, \artiphishell's configuration issue, \atlantis's Java \cc{poi} failure, etc),
and the long tail of real-world edge cases
that automated systems cannot generically handle.
This leads to an interesting observation:
a CRS that reliably applies annotation techniques
in real-world conditions would rank among the top three.
Besides, \atlantis's dominance concentrated in P3--P7
(\autoref{fig:team-performance-per-cpv}),
where many teams underperformed against annotation expectations.

\findingbox{KF 2.}{AIxCC's challenges favor real-world coverage over difficulty. Annotation techniques plus strong system-level solutions could have secured top-three.}

\subsection{PoV Generation Analysis}
\label{s:analysis:pov}

\PP{PF-Solvable CPVs}
PF annotates 34 of 63 CPVs (54\%) as solvable,
but C and Java differ sharply: 30/40 in C (75\%) versus only 4/23 in Java (17\%).
Three factors make Java harder for fuzzing:
\ding{192}~inputs have richer semantic constraints (\eg, XML);
\ding{193}~many timeout/OOM CPVs are fuzzer-unfriendly;
\ding{194}~default OSS-Fuzz seeds for Java are lower quality.
Interestingly, on the C side, \cc{wireshark} (\cc{ws*}), \cc{shadowsocks} (\cc{ss*}),
and \cc{systemd} (\cc{sd*}) contain mostly PF-solvable CPVs by design.
The organizer's challenge notes in these projects indicate they were
intended to evaluate patching and deduplication, not PoV difficulty.
Since they appear in P3--P7, scoring on them required sustained stability.

\PP{CRS over PF}
Six CRSs solved 7--16 CPVs that PF could not (totaling 8/14 in C/Java),
indicating substantial improvement over PF alone.
This advantage can be largely attributed to LLM components,
since LLM-based generation is the most widely developed addition beyond PF,
while non-LLM fuzzing enhancements were explored by only a few teams
(\autoref{s:crs:pov}).
Another direct example: \buttercup, \roboduck, and \fuzzingbrain
built bug-finding stacks consisting mostly of basic parallel fuzzing plus LLM components,
yet all solved multiple CPVs PF missed.
Overall, CRSs exhibit three distinct capabilities.

\BB{Strong targeted detection}
Delta challenges dominate non-PF-solvable CPVs solved by CRSs (15/22).
Given diff-based hints about vulnerability locations,
CRSs analyze code changes and generate triggering inputs directly.
Representative cases include
\cpvd{fp2}{02}, \cpvd{mg1}{01} (C),
and \cpvd{cc1}{03}, \cpvd{po1}{05} (Java).
Specifically, this targeted capability is driven primarily by LLM reasoning:
\atlantis and \bugbuster, the two CRSs that included classic directed fuzzing,
found no uniquely-discovered delta-mode bugs over CRSs without it.
Notably, several such CPVs involve indirect calls
(\cpvd{cu2}{06}, \cpvd{cu4}{03}),
where function pointers obscure the path to vulnerable code
yet CRSs reason through them to reach the bug.

\BB{Overcoming input grammar obstacles}
Some CPVs require inputs conforming to complex grammars
that random mutation cannot satisfy.
Examples include \cc{curl} protocol exploitation
requiring TLV-formatted inputs~\cite{curl-fuzzer} (\cpvd{cu2}{06}),
and structured file formats like PDF with embedded XFA (\cpvf{pb1}{01})
or XLSX for SSRF (\cpvd{po1}{05}).
CRSs can extract format specifications from the source code
and use them to generate syntactically valid inputs.

\BB{Solving logical constraints}
Some CPVs guard vulnerable paths with constraints
that defeat fuzzer feedback mechanisms:
regex patterns (\cpvd{cc1}{03}, \cpvd{po1}{06}),
encoding transformations like URL encoding (\cpvd{cc4}{06}),
Unicode normalization (\cpvd{cc6}{07}, \cpvf{po1}{02}),
or zlib compression (\cpvd{pb1}{02}),
symlink-based path obfuscation (\cpvd{cc7}{08}),
and mathematical guards (\cpvd{cc5}{10}, \cpvd{tk6}{10}).
LLMs can reason about these transformations
and generate constraint-satisfying inputs.

\findingbox{KF 3.}{LLMs complement coverage-based fuzzing in directed (delta-mode) and constraint-heavy contexts.}

\PP{PF-Solvable CPVs Missed by CRSs}
In contrast, these misses stem not from detection capability,
but from trivial yet critical pipeline gaps in handling real-world complexity.

\BB{Broken system dependencies}
Some CPs deviate from standard OSS-Fuzz setups
and break CRS initialization:
\cpvd{av2}{02} uses FuzzTest with its own fuzzer instantiation,
while \cc{pdfbox} CPs ship their own \cc{jazzer} that
overrides the CRS's pre-installed copy.

\BB{Heavy build process}
Unexpected resource demands can exhaust CRS nodes.
Typically, \cc{wireshark} (4.9M LoC)
produces 47 harnesses of $\sim$20.9\,GB each ($\sim$1\,TB total),
causing disk exhaustion or OOM.

\BB{Reproduction behavior mismatch}
CRSs' reproduction criteria can differ subtly from organizer criteria.
Some stateful bugs require multiple executions to trigger:
\cpvf{sd1}{05} crashes on the second execution,
and the organizer runs 100 times by default.
Thus, CRSs verifying with a single execution discarded valid PoVs.

\BB{Incorrect sanitizer}
Some vulnerability classes require specific sanitizers:
\cpvf{da1}{01} (signed integer overflow) needs UBSan~\cite{ubsan}, not ASan~\cite{asan}.
Several teams included UBSan,
yet only \buttercup handled this CPV correctly.

\BB{Crash deduplication granularity}
\cpvf{ss1}{00}--\cpvf{ss1}{04} are distinct heap-buffer-overflows
in \cc{json\_parse\_ex}, differing only by line number.
Coarse-grained deduplication (libfuzzer tokens, function-level grouping, similarity)
failed to distinguish them as separate vulnerabilities.
Engineering can fix these gaps,
but at the cost of domain expertise and careful per-case verification.
This raises an open question that no team has tried:
can LLM integration lower the expertise cost
and generalize robustness more broadly?

\findingbox{KF 4.}{Robust pipeline construction (build, sanitizers, dedup, reproduction) is non-trivial, yet agentic solutions remain under-explored.}

\PP{CPVs Unsolved by Both PF and CRSs}
These CPVs indicate two distinct limits.
\ding{192}~\emph{Reasoning gaps.}
Multi-step cryptographic transformations
(\cpvd{cu3}{07}: XOR/shift; \cpvd{cu4}{08}: AES with Base64 encoding)
accumulate LLM errors end-to-end,
where tool-assisted verification could help.
\ding{193}~\emph{Fuzzing pipeline limitations.}
The \cc{pdfbox} \cc{ExtractTextFuzzer} harness
embeds four timeout and two OOM CPVs (\cpvf{pb1}{03}--\cpvf{pb1}{08})
that expose two coupled weaknesses.
First, timeouts produce near-identical crash signatures,
so deduplication collapses distinct bugs into one.
Second, a shallow timeout, once hit, keeps re-firing
and prevents the fuzzer from reaching deeper bugs:
once \cpvf{pb1}{07} triggers,
repeated timeouts block exploration of \cpvf{pb1}{05} and \cpvf{pb1}{06}.
No CRS attempted on-the-fly patching to bypass shallow timeouts
or fine-grained timeout deduplication,
yielding minimal coverage on this harness.

\findingbox{KF 5.}{Unresolved bugs reveal either reasoning gaps or fuzzing pipeline limitations.}

\subsection{Patch Generation Analysis}
\label{s:analysis:patch}

\PP{MR/CC Patchable CPVs}
MR and CC annotate 31 and 33 of 63 CPVs as patchable, respectively,
with 36 unique CPVs covered in total.
Although more than half of all CPVs are covered,
this does not indicate superior patching capability;
rather, it reflects that MR and CC can solve
basic fix challenges
where the root cause is directly surfaced by the sanitizer report
and the local code context suffices to derive a fix
without cross-reference reasoning:
textbook vulnerabilities with well-known fix patterns.
Typical examples include
boundary checks for buffer overflows
(\cpvd{fp2}{02}, \cpvf{ss1}{00}--\cpvf{ss1}{04}),
secure XML parser configuration for XML External Entity (XXE) (\cpvf{pb1}{00}, \cpvf{pb1}{01}),
decode-then-normalize reordering for path traversal
(\cpvd{cc4}{06}, \cpvd{cc6}{07}),
single-line fixes such as null checks
and format-string corrections
(\cpvf{cm1}{01}, \cpvd{cu5}{01}),
and CTF-style backdoors labelled by a ``flag'' string
that simply need deletion
(\cpvd{cu2}{06}, \cpvf{sd1}{03}, \cpvd{cc5}{10}).

\PP{Semantically Incorrect Patches from MR/CC}
A notable fact is that
a significant fraction of generated patches
pass all automatic validation, yet
contain semantic issues caught only by manual review
(CC: 20/53, 37.7\%, MR: 26/57, 45.6\%).
\BB{Wrong root cause}
Agents may suppress the crash symptom
rather than addressing the underlying defect.
One typical pattern is defensive patching in parsing logic, when root cause and crash location are distant (\cpvd{av2}{02}, \cpvf{sd1}{05}, \cpvf{pb1}{03}, \cpvf{pb1}{04}).
For timeout CPVs (\cpvf{pb1}{03}, \cpvf{pb1}{04}, \cpvd{po1}{06}, \cpvf{pb1}{08}),
a common error is inserting a hard-coded iteration limit
instead of fixing the infinite loop root cause.

\BB{Incomplete fix}
Agents can fix the specific crashing path
instead of completely remediating the underlying bug.
In \cc{mongoose} (\cpvf{mg1}{00}, \cpvd{mg1}{01}, \cpvd{mg2}{02}),
\cc{\#line} directives hide the amalgamated compilation unit
from the sanitizer reports,
so agents patched the reported files but missed it.
Other cases include
the XXE-parser hardening of \cpvf{pb1}{00}
that omits a critical security setting,
and the dangling-pointer cases (\cpvf{sd1}{05}, \cpvd{ex2}{01})
where patches go at the call site
instead of inside the buggy memory API.

\BB{Functionality deviation}
Some patches eliminate the crash
but subtly alter program semantics
in ways that functional tests do not cover
(\eg, \cpvf{ws2}{04}, \cpvf{ws1}{02}, \cpvf{ws1}{05}, \cpvf{ws1}{11},
\cpvd{lx3}{04}, \cpvf{sd1}{01}, \cpvd{tk6}{10}, \cpvd{fp3}{03}).
In \cpvd{av2}{02},
a patch produces incorrect edge colors
because YUV-to-RGB conversion depends on neighboring pixels
that the patch mishandles.
In \cpvd{cc7}{08},
a patch resolves relative symbolic links to absolute paths
during archive extraction,
when it should only validate against path traversal
without modifying the link target.

\BB{Introducing new bugs}
In \cpvf{ss1}{00},
an extra \cc{break} renders
part of the encoding logic unreachable;
in \cpvf{sd1}{05},
removing a \cc{mfree} call introduces a memory leak.

\BB{Missing domain knowledge}
In \cpvd{ws7}{13},
correct patching requires understanding
how GVCP~\cite{gvcp} bootstrap registers,
which are standardized memory-mapped addresses for device discovery,
are managed within the project.
In \cpvf{ws1}{01},
patches confused self-reported packet lengths with verified ones,
missing the knowledge to tell them apart.

\findingbox{KF 6.}{Claude Code slightly outperforms MultiRetrieval, but both suffer 38--46\% semantic incorrectness.}

\PP{CRS over MR/CC}
On one hand,
CRSs solved 16 cases where MR/CC failed,
demonstrating capabilities beyond foundational agents,
such as the correct fixes
for incomplete remediations (\cpvf{sd1}{05}, \cpvd{ex2}{01}),
functionality deviations
(\cpvf{ws1}{11}, \cpvd{lx3}{04}, \cpvf{sd1}{01}, \cpvd{tk6}{10}),
and misleading sanitizer reports
(\cpvd{mg2}{02}, \cpvf{mg1}{00})
listed in the previous paragraph.
On the other hand,
CRSs also suffer from semantically incorrect patches,
though mostly at lower rates than MR/CC (\autoref{t:game-strategy}).
\atlantis and \buttercup achieve 83.8\% and 79.2\% patch accuracy, respectively;
\artiphishell reaches 100\%,
though this likely reflects strict patch filtering
(\eg, its 5-minute post-patch fuzzing step)
and patched challenge distribution
rather than full mitigation of semantic correctness.
In contrast,
\roboduck and \fuzzingbrain fall to 31.7\% and 23.3\% accuracy,
concentrated in bursts of invalid submissions
on one or two challenges, likely
tied to strategy issues
such as parallel generation without adequate deduplication.
Despite these differences,
a common factor among higher-accuracy CRSs
is the adoption of multi-PoV validation, post-patch fuzzing,
or LLM-based reflection (\autoref{s:crs:patch}),
yet incorrect patches remain prevalent overall.

\findingbox{KF 7.}{High-accuracy CRSs can achieve 16--21\% semantic incorrectness, substantially below MultiRetrieval/ Claude Code's 38--46\% but still non-negligible.}

\PP{MR/CC Patchable CPVs Failed by CRSs}
Most CRSs incorporate comparable patch agents,
so these 9 failures likely stem not from capability gaps
but from other factors:
missing PoVs that never triggered patch generation,
scheduling pressure from too many concurrent PoVs,
and system-wide stability issues discussed earlier.
Interestingly, one identifiable cause is
\emph{deployment-time configuration trade-offs}.
For example, \cpvd{ws5}{08} is locally patchable by MR (one agent in \atlantis),
but \atlantis's 30-minute per-CPV timeout,
necessary to manage dozens of concurrent challenges,
is largely consumed by the \cc{wireshark} build alone,
leaving insufficient time for the patch loop to complete.

\findingbox{KF 8.}{Pipeline construction challenges and patch-strategy tradeoffs can limit CRSs' performance.}

\PP{CPVs Failed by both CRSs and MR/CC}
Seven Java CPVs were never patched:
4 infinite loops (\cpvf{pb1}{03}, \cpvf{pb1}{04}, \cpvf{pb1}{08}, \cpvf{po1}{00}),
1 integer overflow (\cpvf{pb1}{06}),
1 ReDoS (\cpvd{po1}{06}),
and 1 JVM crash via obfuscated backdoor (\cpvf{po1}{03}).
These cases reveal two limitations of current patching.
\ding{192}~\emph{Reasoning gaps.}
For example, ReDoS (\cpvd{po1}{06}) requires regex worst-case reasoning
that agents cannot perform reliably end-to-end;
integrating non-LLM regex analysis tools such as symbolic regex repair~\cite{li:vulcanboost}
could help.
\ding{193}~\emph{Patch pipeline limitations.}
Current patch pipelines rely heavily on crash stack traces
from the PoV/fuzzing side for root cause analysis,
which degrades sharply when such signals are absent.
The \cc{pdfbox} infinite-loop CPVs and \cpvf{po1}{03} demonstrate this:
timeouts and JVM-level crashes give only minimal localization,
and most CRSs failed on these cases.
Approaches that go beyond this limitation remain to be explored.

\findingbox{KF 9.}{As with bug finding, unresolved patches face either reasoning gaps or patch pipeline limitations.}

\PP{Noteworthy No-PoV Patches in shadowsocks}
In \cc{shadowsocks} (\cpvf{ss1}{00}--\cpvf{ss1}{04}),
three CRSs patched CPVs for which they had no PoV,
even though two of them require PoVs to generate patches.
This was possible because all five heap-buffer-overflows
represent the same bug pattern repeated at different locations
within a large JSON parsing function;
some CRSs recognized the pattern and generated patches
that fixed all instances together.

\subsection{Other Analyses}
\label{s:analysis:other}

\PP{SARIF Validation}
\autoref{fig:team-performance-per-cpv}~(v) presents
SARIF validation results across 13 broadcasts
(8 valid, 5 invalid).
PoV-centric teams (\autoref{s:crs:sarif})
can only submit \emph{Correct} when a PoV matches; thus,
they were unable to assess the 5 invalid reports,
leading to fewer submissions
(\atlantis: 8, \fuzzingbrain: 7, \buttercup: 1).
With PoV evidence,
\fuzzingbrain and \buttercup achieved 100\% accuracy,
but \atlantis falsely matched PoVs to 2 invalid reports.
In contrast, non-PoV-centric teams
can assess all broadcasts but risk wrong answers:
\bugbuster scored 10/13 and \artiphishell 9/13.

\PP{Bundle Results}
\input{tbl/bundle-composition}
\autoref{t:bundle-results} shows bundle outcomes.
Overall accuracy is high (91/99, 92\%):
all seven teams adopted
PoV-based patch generation (\autoref{s:crs:game})
that naturally pairs PoVs with their patches,
accounting for 86\% of bundles with 93\% accuracy.
Patch-SARIF, the only non-PoV pairing,
achieved 1/3 accuracy.
Of the 8 incorrect bundles,
only one is a true pairing mismatch;
the other 7 failed due to unsuccessful patches,
confirming patch quality as the practical bottleneck.

\PP{Resource Usage and Efficiency}
No team consumed the full quota of either resource;
\autoref{t:resource-utilization} reports per-team breakdowns.
\ding{192}~LLM spending concentrates in two providers:
$\sim$94\% on Anthropic and OpenAI,
$\sim$6\% on Gemini and xAI.
\ding{193}~LLM spend rank closely tracks the final score,
with only \fuzzingbrain and \roboduck flipping the order.
\ding{194}~From a cost-efficiency view (score per \$K total spend),
\roboduck, \buttercup, and \atlantis lead.

\PP{0-Day Discovery}
All seven teams discovered at least one 0-day,
yielding 25 distinct vulnerabilities
across 10 OSS projects,
of which 12 (48\%) were patched
(\autoref{fig:team-performance-per-cpv}~(ii, iv)).
See \cite{sok-website} for more details.
Responsible disclosure was coordinated
by Kudu Dynamics with OSTIF and ADALogics.

%% file: tbl/competition-results.tex
\begin{figure}[t]
\centering
\includegraphics[width=\columnwidth]{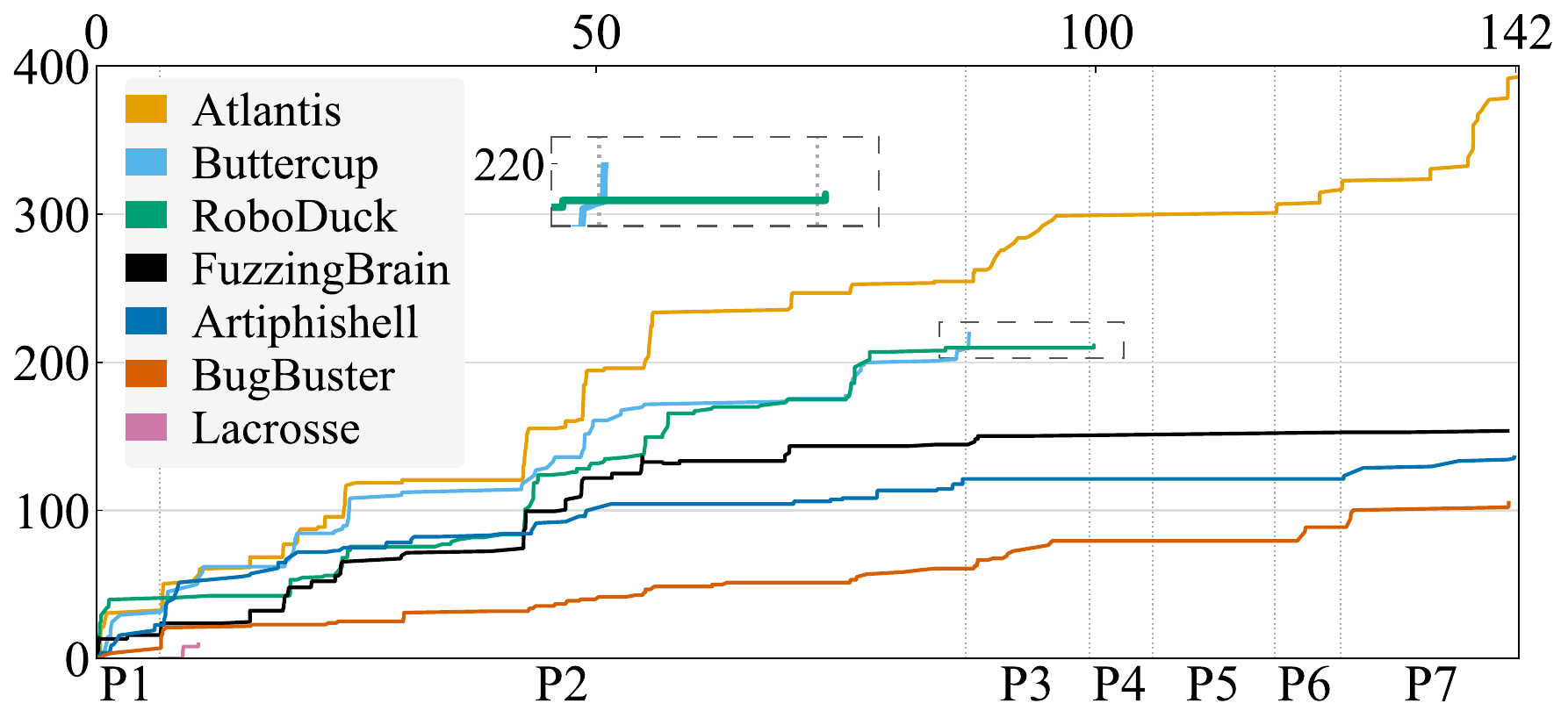}
\caption{Score per time (top) and phase (bottom) axes.}
\label{f:score-over-time}
\end{figure}

%% file: tbl/score-breakdown.tex
\begin{table}[t]
\centering
\caption{AFC Score Breakdown.
Columns are ordered left-to-right by final score (descending).
Pen.\ stands for penalties.}
\label{t:score-breakdown}
\input{tbl/competition-results-simple}
\end{table}

%% file: tbl/competition-results-simple.tex
{\rowcolors{2}{}{gray!15}
\smaller
\setlength{\tabcolsep}{4pt}
\adjustbox{trim=0pt 0pt 3pt 0pt,clip,max width=\columnwidth}{
\begin{tabular}{@{}>{\cellcolor{white}}l l ccccccc@{}}
\toprule
& & \hyperlink{team:AT}{\texttt{AT}} & \hyperlink{team:TB}{\texttt{TB}} & \hyperlink{team:TI}{\texttt{TI}} & \hyperlink{team:FB}{\texttt{FB}} & \hyperlink{team:SP}{\texttt{SP}} & \hyperlink{team:42}{\texttt{42}} & \hyperlink{team:LC}{\texttt{LC}} \\
\midrule
& C & \underline{\textbf{52.6}} & 31.0 & 22.7 & 22.6 & 31.4 & 49.1 & 1.5 \\
& Java & 27.0 & 21.3 & \underline{\textbf{31.6}} & 29.7 & 16.5 & 21.1 & 0.0 \\
\multirow{-3}{*}{\textbf{PoV}} & Sum & \underline{\textbf{79.6}} & 52.4 & 54.3 & 52.3 & 47.8 & 70.1 & 1.5 \\
\midrule
& C & \underline{\textbf{113.5}} & 74.2 & 51.8 & 45.3 & 40.5 & 9.7 & 4.9 \\
& Java & \underline{\textbf{57.5}} & 26.7 & 49.8 & 23.5 & 13.8 & 4.5 & 0.0 \\
\multirow{-3}{*}{\textbf{Patch}} & Sum & \underline{\textbf{171.0}} & 100.9 & 101.6 & 68.8 & 54.3 & 14.2 & 4.9 \\
\midrule
& C & 5.0 & 1.0 & 3.9 & 4.7 & 7.5 & \underline{\textbf{8.7}} & 0.0 \\
& Java & 1.0 & 0.0 & 1.0 & \underline{\textbf{1.5}} & 1.0 & 1.0 & 0.0 \\
\multirow{-3}{*}{\textbf{SARIF}} & Sum & 6.0 & 1.0 & 4.9 & 6.2 & 8.5 & \underline{\textbf{9.7}} & 0.0 \\
\midrule
& C & \underline{\textbf{99.6}} & 49.4 & 26.0 & 27.6 & 18.3 & 7.8 & 3.2 \\
& Java & \underline{\textbf{36.6}} & 15.7 & 23.8 & -1.1 & 7.0 & 3.2 & 0.0 \\
\multirow{-3}{*}{\textbf{Bundle}} & Sum & \underline{\textbf{136.2}} & 65.1 & 49.8 & 26.4 & 25.3 & 11.0 & 3.2 \\
\midrule
& C & \underline{\textbf{270.6}} & 155.6 & 104.4 & 100.2 & 97.6 & 75.4 & 9.6 \\
& Java & \underline{\textbf{122.1}} & 63.8 & 106.3 & 53.5 & 38.3 & 29.7 & 0.0 \\
& Pen. & -0.4 & -0.6 & -16.3 & -13.5 & \underline{\textbf{-0.1}} & -0.3 & -1.1 \\
\multirow{-4}{*}{\textbf{Total}} & Final & \underline{\textbf{392.8}} & 219.4 & 210.7 & 153.7 & 135.9 & 105.0 & 9.6 \\
\bottomrule
\end{tabular}
}
}

%% file: tbl/bundle-composition.tex
\begin{table}[t]
\centering
\caption{Bundle strategies and results.
Team abbr: \autoref{t:crs-philosophy}.}
\label{t:bundle-results}
{\rowcolors{2}{}{gray!15}
\smaller
\setlength{\tabcolsep}{4pt}
\adjustbox{trim=0pt 0pt 0pt 0pt,clip,max width=\columnwidth}{
\begin{tabular}{@{}l ccccccc | c@{}}
\toprule
& \hyperlink{team:AT}{\texttt{AT}} & \hyperlink{team:TB}{\texttt{TB}} & \hyperlink{team:TI}{\texttt{TI}} & \hyperlink{team:FB}{\texttt{FB}} & \hyperlink{team:SP}{\texttt{SP}} & \hyperlink{team:42}{\texttt{42}} & \hyperlink{team:LC}{\texttt{LC}} & Total \\
\midrule
\textbf{PoV-Patch}  & 27/28 & 18/18 & 16/18 & 6/9 & 7/7 & 4/4 & 1/1 & 79/85 \\
\textbf{PoV-SARIF}  & 1/1 & -- & -- & -- & -- & -- & -- & 1/1 \\
\textbf{Patch-SARIF}  & -- & -- & 1/1 & 0/2 & -- & -- & -- & 1/3 \\
\textbf{PoV-Patch-SARIF} & 7/7 & 1/1 & -- & 2/2 & -- & -- & -- & 10/10 \\
\midrule
\textbf{Accuracy} & 35/36 & 19/19 & 17/19 & 8/13 & 7/7 & 4/4 & 1/1 & 91/99 \\
\textbf{Score} & 136.2 & 65.1 & 49.8 & 26.4 & 25.3 & 11.0 & 3.2 & 316.9 \\
\bottomrule
\end{tabular}
}}
\end{table}

%% file: lessons.tex
\section{Lessons and Future Directions}
\label{s:lesson}

\PP{From Competition to Industry Deployment}
Reflecting on the competition,
we identify several areas where future efforts
could further ease the path to practical deployment.

In the final, CRSs operated on self-provisioned Azure clusters
with budgets of hundreds of dollars per challenge,
ensuring resource availability would not limit technical exploration.
This contrasts with individual developers or small teams
who need lightweight, single-machine solutions at minimal cost.
Although finalist CRSs have been open-sourced,
their resource usage models and runtime environments
differ substantially from typical deployment settings,
posing barriers to post-competition adoption.
Future work could develop resource-efficient CRS variants
that remain effective under constrained environments;
future competitions could also introduce resource-limited tracks
that account for the needs of individual developers and small teams.

Beyond resource constraints,
OSS communities need time to adapt their pipelines for CRS integration,
such as provisioning LLM services,
standardizing CRS interfaces for broader OSS applications,
defining end-to-end workflows from AI-assisted bug finding to patch submission,
and ensembling multiple CRSs for combined effectiveness.
Considering these needs at the design stage
would lower adoption barriers for OSS maintainers and practitioners.
On the post-competition side,
initiatives like OSS-CRS~\cite{oss-crs} have begun
to address these needs;
future competitions could learn from this
by co-designing deployment pathways with OSS communities
for smoother transition to real-world adoption.

\PP{From Competition to Research Advancement}
The competition design and team-built systems
hold significant research value,
yet certain design improvements could be made to further enlarge their value as research assets.

AIxCC is intrinsically a substantial experimental investment,
yet its telemetry primarily serves real-time monitoring and scoring
rather than retrospective analysis of \emph{why} systems behaved as they did.
Designing telemetry with post-hoc analysis as a first-class goal---logging
intermediate outputs, decision traces, and environmental snapshots---would
enable systematic studies of failure modes and technique effectiveness.
Organizer-built baseline CRSs participating alongside teams
would further enrich such analysis
by providing reference points for comparison.

The competition's exploration of open-source LLMs was minimal:
although two teams fine-tuned open-source models,
only one ultimately deployed them.
The competition structure offered little incentive to invest in open-source models,
as the uncertainty, cost, and data-acquisition difficulty of fine-tuning
made prompt-based techniques on frontier commercial models
a more predictable and cost-effective strategy.
While advancing frontier AI for cybersecurity is a natural focus,
open-source models offer distinct value to the OSS community
through lower cost, customizability, and transparency.
A dedicated sub-track comparing CRS performance
under open-source models would encourage exploration in this direction.

\PP{Areas of Expansion}
As the first large-scale competition of its kind,
AIxCC necessarily scoped its focus.
Several directions not covered in this iteration
are worth exploring in future editions:
\emph{full autonomy} (generating harnesses and handling arbitrary build systems),
\emph{multi-CRS settings} (collaborative analysis or adversarial formats
where CRSs attack competitors' patches),
and \emph{semantic correctness evaluation} (approaches for patch semantic correctness).

\PP{AI-Powered Vulnerability Scanning}
Early 2026 has seen notable advances in
frontier LLMs' cybersecurity capabilities,
with works like Claude Mythos Preview~\cite{claude-mythos}
and coding agent patch evaluation~\cite{ta-patch-2026}
showing striking performance in bug finding and repair.
The capability jump signals further rapid progress
in both offensive and defensive automation,
which is exactly the future AIxCC was set up to prepare for.
Technically, AIxCC's exploration and this LLM capability surge
are complementary and mutually beneficial.
On one hand, pure-LLM pipelines still carry fundamental limitations
such as hallucination and nondeterminism,
so the LLM--non-LLM cooperation patterns established in AIxCC
remain a direct and meaningful reference
for CRSs built on top of stronger base models.
On the other hand, stronger LLMs also open room
for the next generation of CRSs to simplify their architectures
and push autonomy further.

%% file: validity.tex
\section{Limitations}
\label{s:limitations}

Our study has three notable limitations.
\ding{192}~\emph{Taxonomy interpretation.}
We strictly follow enabled functionalities in submission-version code,
treating code as authoritative when it conflicts
with questionnaire or meeting records;
two security experts cross-validated each profile,
and the final taxonomy was shared with all seven teams
(three had bandwidth and confirmed).
Still, any misreading of the source would propagate
as taxonomy misinterpretation.
\ding{193}~\emph{Annotation accuracy.}
PF/MR/CC annotation is an approximation
under estimated resources and selected techniques.
On the PoV side,
longer fuzzing, a different fuzzer framework,
or some teams' private initial corpora
could each mark additional CPVs as solvable;
the annotation is thus an approximation of what this
representative configuration would solve in AIxCC,
not a general claim about fuzzing capability.
We run each technique three times and take the union
as a stable lower bound.
\ding{194}~\emph{No ablation study.}
The most accurate way to understand the techniques used by these teams
would be to migrate them into a single platform
and measure their contributions under controlled resources.
We believe this is out of scope for a SoK
given the resource cost (CRSs use LLMs heavily)
and engineering cost (CRSs are large, heterogeneous systems).
Post-competition efforts such as OSS-CRS~\cite{oss-crs}
are working on this direction.

%% file: conclusion.tex
\section{Conclusion}
\label{s:conclusion}

AIxCC represents a milestone in autonomous cybersecurity research,
demonstrating that AI-powered CRSs can discover and patch vulnerabilities
in real-world software at scale.
Through systematic analysis of competition design, CRS architectures, and results,
we summarized technical insights of those systems,
revealed both their genuine performance advances
and the persistent gap between technique capability and system reliability.
We hope this work serves as a foundation
for future competition designs, CRS development, and practical deployment
of autonomous cybersecurity systems.

%% file: ack.tex
\section*{Acknowledgments}
\label{s:ack}

This work is a joint effort of
multiple teams and organizations involving dozens of contributors,
who collectively aim to provide
a systematic and insightful view of the AIxCC competition.
We are grateful to every one of them for making this work possible
and highlight their primary contributions.

This work was initiated and directed by Taesoo Kim,
who supervised the entire research process.
Cen Zhang led the structure design,
the CRS code study and team meetings,
cross-team collaboration,
coordination with contributors
to distill the core findings of each part,
and the paper drafting.
Younggi Park implemented the patch analysis framework,
conducted the experiments,
manually validated agent-generated patches,
and helped analyze and draft the patch analysis findings.
Fabian Fleischer verified and integrated competition data,
visualized the results
(such as \autoref{fig:team-performance-per-cpv}),
helped validate patches and drafted patch analysis,
and prepared the artifact.
Yu-Fu Fu compiled team statistics,
studied and summarized CRS patch systems of all teams,
and helped draft the patch techniques.
Jiho Kim conducted PoV generation analysis
in \autoref{s:analysis:pov}
and contributed to the discussions in \autoref{s:lesson}.
Dongkwan Kim contributed in team meeting discussions,
conducted the parallel fuzzing experiments,
authored \autoref{s:analysis:other},
studied teams' bundling strategies,
and helped draft \autoref{s:crs:game}.
Youngjoon Kim studied CRS SARIF techniques of all teams and some teams' CRSs
during the early code study,
authored \autoref{s:cp},
and helped draft \autoref{s:crs:sarif}.
Qingxiao Xu analyzed
the inaccurate patch submissions across teams,
and both Qingxiao Xu and Ze Sheng
assisted with the analysis in \autoref{s:analysis:other}.
Andrew Chin studied some CRSs' bug finding techniques
during the early code study,
cross-validated the technique taxonomy in \autoref{s:crs},
and helped check and refine the paper during the final stages.
The above Team Atlanta authors, along with Hanqing Zhao,
authors from Team Fuzzing Brain (Jeff Huang, Ze Sheng, and Qingxiao Xu),
and Team Lacrosse (Michael Pelican, David J.\ Musliner),
cross-validated the technique taxonomy in \autoref{s:crs}
and conducted proofreading.
We also thank external contributors:
Joshua Wang for help with taxonomy cross-validation
and proofreading,
and Brian J.\ Lee for help preparing the artifact.

We sincerely thank the remaining four AIxCC teams---Trail of Bits (Michael Brown),
Theori (Tyler Nighswander),
Shellphish (Wil Gibbs and Yan Shoshitaishvili),
and 42-b3yond-6ug (Xinyu Xing)---for
their support, feedback, discussions, and meetings,
as well as the internal materials they shared
(\emph{e.g.}, Wil Gibbs shared Shellphish's internal white paper),
which helped shape our understanding
of each CRS's overall design motivations.
Xinyu Xing also shared his insights and reflections
on the competition,
such as the gap between research and practice,
during early discussions of this work.

Kudu Dynamics, as the competition organizer,
provided raw materials on competition design
from the organizer's perspective,
shared competition data
and continually updated it as our analysis needs evolved,
and held weekly meetings for communication and coordination.
Nicholas Vidovich and Matthew Lehman led
the organizer collaboration effort
and contributed organizer-side content and review.
Isaac Goldthwaite contributed content on competition rules, challenges, and scoring.
Jefferson Casavant contributed content on competition infrastructure and implementation.
Jon Silliman and Mikel Mcdaniel ensured
the paper statistics and data aligned with
the competition dataset and results.
We also thank DARPA for its generous support and prompt responses to our inquiries,
including granting full access to
the challenge source code and competition data.

This research was supported by
the Advanced Research Projects Agency for Health (ARPA-H)
under Other Transaction Agreement No.\ 140D042590046,
and by a gift from Team Atlanta.

%% file: ethics.tex
\section*{Ethical Considerations}
\label{s:ethics}

\PP{Stakeholder Identification}
We identify three primary stakeholder groups.
\emph{(1)~Researchers and practitioners}:
the seven finalist teams,
competition organizers
(DARPA, ARPA-H, and Kudu Dynamics),
and security researchers
who may build upon our findings,
whose system designs, performance data,
and strategic decisions are analyzed in detail.
\emph{(2)~Open-source community}:
developers and maintainers
of the 24 open-source projects
from which challenge projects were derived,
as well as the broader OSS ecosystem
that depends on the security of these projects.
\emph{(3)~LLM vendors}:
whose models were used by competing teams
and whose API usage patterns are discussed.

\PP{Ethical Principles}
\emph{Beneficence.}
This work advances the understanding
of autonomous Cyber Reasoning Systems
and their application to real-world vulnerability detection.
By systematically analyzing competition design,
CRS architectures, and performance outcomes,
we provide actionable insights
for the OSS security community,
future competition organizers,
and researchers building autonomous security systems.

\emph{Respect for Persons.}
We had discussions
with organizers and competing teams
to gather firsthand accounts of design decisions
and operational experiences,
and maintained ongoing communication throughout the writing process
to ensure accurate representation of their work.
The AIxCC competition rules
permitted publication of CRS source code,
performance data, and architectural details;
all finalist teams were aware that their systems and results
would be analyzed,
and were given draft sections for review (3 confirmed).

\emph{Justice.}
All seven finalist teams
are analyzed with equal rigor
and presented with consistent methodology.
No team is singled out or unfairly characterized.
We did not re-evaluate any team's CRS independently;
all performance data originates from the final competition results.
The organizers have no conflict of interest
with any of the seven finalist teams
and conducted fair evaluation
through extensive communication and documentation.
Competition data will be publicly released
to ensure transparency and reproducibility,
subject to DARPA's approval and disclosure guidelines.

\emph{Respect for Law and Public Interest.}
The AIxCC competition operated
under DARPA and ARPA-H research frameworks,
with all teams agreeing to rules
governing data handling and disclosure.
0-day vulnerabilities were reported
through responsible disclosure processes
in compliance with applicable laws.
We document our methodology and data sources
to enable reproducibility;
competition data will be publicly released
to ensure transparency and accountability.

\PP{Potential Harms}
This work does not involve human subjects or private user data.
Discussions with teams focused on technical methodology
and system architecture,
not personal or sensitive information.
We identify three potential tangible harms.
First,
\emph{misinterpretation of performance data}
could cause reputational or financial harm to specific teams
if rankings or analyses are taken out of context.
Second,
\emph{0-day vulnerability exposure}
could be exploited by malicious actors
before patches are available.
Third,
\emph{dual-use concerns} arise on multiple paths:
adapting CRS architectures for automated exploit generation,
leveraging our taxonomy to accelerate offensive tool development,
and misusing competition data
(particularly team-found 0-day PoCs)
for real-world attacks before patches reach end users.

\PP{Mitigations}
We took the following steps to address these risks.
For \emph{misinterpretation of performance data},
each section was cross-validated by at least two authors.
We also had discussions with all seven finalist teams
and maintained regular communication throughout the writing process
to ensure accurate representation.
For \emph{0-day vulnerability exposure},
all 0-day vulnerability data
originates exclusively from the final competition environment.
Disclosure began shortly after the August 2025 final
and remains ongoing through standard responsible disclosure protocols:
Kudu Dynamics, OSTIF, and ADALogics
coordinate with each affected upstream maintainer via private channels
(typically email or security mailing lists),
share PoVs and remediation guidance,
and embargo public details until fixes are released
or a standard 90-day window elapses.
All discovered 0-day vulnerabilities
have been reported through this process;
detailed 0-day information
is not included in this paper or its artifacts.
For \emph{dual-use concerns},
CRS performs not only vulnerability detection
but also vulnerability repair.
This aligns with the philosophy of OSS-Fuzz:
by enabling defenders to find and patch vulnerabilities
faster than malicious actors can exploit them,
CRS contributes positively to defense.

\PP{Team Well-being}
AIxCC spanned two years,
demanding long-term commitment across
both cutting-edge research and production-grade engineering,
with inherent burnout risks.
Evolving competition rules and rapidly advancing AI capabilities
required teams to continuously adapt,
adding further pressure.
Some teams' final outcomes were undermined
by system instability rather than capability gaps,
which can be particularly frustrating after such sustained investment.
For this study,
all participating teams consented to questionnaires
in flexible formats depending on their availability,
and draft sections were shared with all seven teams for review.

\PP{Decision to Conduct and Publish}
\emph{Decision to research.}
AIxCC represents the largest competition to date
for LLM-based autonomous vulnerability detection and repair,
yet no systematic analysis of its design,
CRS approaches, or outcomes existed prior to this work.
We determined that the research was justified
by the need to document lessons learned,
identify genuine technical advances,
and surface open challenges
for security research.

\emph{Decision to publish.}
The decision to publish was made
after communicating with all participating teams,
competition organizers, and sponsoring agencies.
We believe our insights on CRS design and implementation
will benefit future CRS developers
and software security researchers.
Since we also provide insights into vulnerability repair techniques,
the security benefits outweigh potential dual-use risks.

%% file: openscience.tex
\section*{Open Science}
\label{s:openscience}

All data, scripts, finalist team questionnaires, and meeting notes
are archived in our artifact~\cite{sok-artifact},
with a companion website~\cite{sok-website}
indexing the artifact, our extended analysis, public finalist documentation, and official challenge set access links;
competition data and the analysis framework
await DARPA's official release.

%% file: app.tex
\section{Per-Team Details}
\label{s:app:extended}

\PP{Resource Usage}
\autoref{t:resource-utilization} reports per-team consumption
of Azure compute and LLM API budgets.

\input{tbl/resource-utilization}

\PP{Submission Accuracy}
\label{s:app:submission-accuracy}
\autoref{t:game-strategy} reports per-team counted submissions
(correct and incorrect, excluding duplicates)
and accuracy rates by scoring category.

\input{tbl/game-strategy}

\PP{PoV Generation Techniques}
\label{s:app:pov-techniques}
\autoref{t:fuzzing-techniques} extends \autoref{t:pov-overview}
with per-team tool and parameter details.
Teams generally fall into the structure of \autoref{s:crs:pov},
but apply language- and challenge-mode-specific adaptations.
Java's logical vulnerabilities often stem from unsafe sink function usage,
prompting teams to adopt sink-targeted PoV generation,
directed fuzzing toward sinks, and improved sanitizers.
For delta-mode challenges,
teams narrow scope from the full codebase to the diff and related code for more targeted bug detection.
For SARIF broadcasts received mid-competition,
teams treat them as pre-specified bug candidates
and use them to guide the PoV generation.

\input{tbl/fuzzing-techniques}

\PP{Patch Generation Techniques}
\label{s:app:patch-techniques}
\autoref{t:crs-patch-techniques} extends \autoref{t:patch-overview}
with per-team tool and configuration details.
Teams generally fall into the structure of \autoref{s:crs:patch},
adapting to challenge types by adjusting language-specific tooling and prompts,
along with CWE- or bug-type-specific guidance.
For delta mode,
most CRSs focus vulnerability analysis on modified code from diff files.
Additionally, \roboduck runs two diff-analysis agents in parallel,
one filtering out compiler-unused files and one using the complete diff,
to broaden analysis from immediate change sites
to all affected code.

\input{tbl/patch-techniques}

\section{Abbreviations}
\label{s:app:abbr}

\PP{CP/CPV naming rule}
A CP is identified as \cc{<project><cp-idx><mode>},
and a CPV as \cc{<project><cp-idx><mode><cpv-idx>}, where
\cc{<project>} follows the \emph{Abbr.} column of \autoref{t:cp-overview};
\cc{<mode>} is \cc{$\square$} for full-mode or \cc{$\blacktriangle$} for delta-mode;
\cc{<cp-idx>} and \cc{<cpv-idx>} are numeric indices within the project and CP, respectively.

\PP{Cheat sheet}
The following shows acronyms used in paper.

\vspace{0.6em}
\noindent
\setlength{\fboxsep}{6pt}%
\colorbox{black!5}{%
\begin{minipage}{\dimexpr\columnwidth-2\fboxsep\relax}
\footnotesize
\setlength{\tabcolsep}{3pt}
\renewcommand{\arraystretch}{1.05}
\begin{tabularx}{\linewidth}{@{}l X @{\hskip 0.6em} l X@{}}
\textbf{42}  & \Lbugbuster              & \textbf{MR}    & MultiRetrieval \\
\textbf{AFC} & AIxCC Final Competition  & \textbf{PF}    & Parallel Fuzzing \\
\textbf{ASC} & AIxCC Semifinal Comp.    & \textbf{PoV}   & Proof of Vulnerability \\
\textbf{AT}  & \Latlantis               & \textbf{RCA}   & Root Cause Analysis \\
\textbf{CC}  & Claude Code              & \textbf{SAST}  & Static Application \\
\textbf{CP}  & Challenge Project        &       & \quad Security Testing \\
\textbf{CPV} & Challenge Project Vuln.  & \textbf{SARIF} & Static Analysis Results \\
\textbf{CRS} & Cyber Reasoning System   &       & \quad Interchange Format \\
\textbf{CWE} & Common Weakness Enum.    & \textbf{SP}    & \Lartiphishell \\
\textbf{FB}  & \Lfuzzingbrain           & \textbf{TB}    & \Lbuttercup \\
\textbf{LC}  & \Llacrosse               & \textbf{TI}    & \Lroboduck \\
\end{tabularx}
\end{minipage}%
}

\clearpage

%% file: tbl/resource-utilization.tex
\begin{table}[!ht]
\centering
\caption{Resource Utilization.
Team abbr: \autoref{t:crs-philosophy}.}
\label{t:resource-utilization}
\smaller
\setlength{\tabcolsep}{5pt}
\adjustbox{trim=0pt 0pt 0pt 0pt,max width=\columnwidth,clip}{
\rowcolors{2}{}{gray!15}
\begin{tabular}{@{}l rrrrrrr@{}}
  \toprule
  & \hyperlink{team:AT}{\texttt{AT}} & \hyperlink{team:TB}{\texttt{TB}} & \hyperlink{team:TI}{\texttt{TI}} & \hyperlink{team:FB}{\texttt{FB}} & \hyperlink{team:SP}{\texttt{SP}} & \hyperlink{team:42}{\texttt{42}} & \hyperlink{team:LC}{\texttt{LC}} \\
  \midrule
  Azure (\$K) & 73.9 & 18.5 & 20.3 & 63.2 & 54.9 & 38.7 & 7.1 \\
  \midrule
  OpenAI (\$K) & 6.6 & 0.6 & 8.2 & 3.1 & 0.4 & 0.4 & 0.4 \\
  Anthropic (\$K) & 20.0 & 20.6 & 3.1 & 7.5 & 2.6 & 0.7 & 0.3 \\
  Gemini (\$K) & 2.8 & -- & 0.2 & 1.6 & -- & -- & {$<$0.1} \\
  xAI (\$K) & {$<$0.1} & -- & -- & -- & -- & -- & -- \\
  All LLMs (\$K) & 29.4 & 21.1 & 11.5 & 12.2 & 2.9 & 1.1 & 0.6 \\
  \midrule
  Total (\$K) & 103.3 & 39.6 & 31.8 & 75.4 & 57.8 & 39.8 & 7.8 \\
  Score / \$K & 3.80 & 5.54 & 6.63 & 2.04 & 2.35 & 2.64 & 1.25 \\
  \bottomrule
\end{tabular}
}
\end{table}

%% file: tbl/game-strategy.tex
\begin{table}[!ht]
\centering
\caption{Submission statistics.}
\label{t:game-strategy}
\scriptsize
\setlength{\tabcolsep}{3pt}
\resizebox{\columnwidth}{!}{
\rowcolors{3}{gray!15}{}
\begin{tabular}{@{}l rrrr rrrr r@{}}
\toprule
 & \multicolumn{4}{c}{\textbf{Counted Submissions}} & \multicolumn{4}{c}{\textbf{Accuracy (\%)}} & \textbf{AM} \\
\cmidrule(lr){2-5} \cmidrule(lr){6-9}
\textbf{Team} & \textbf{PoV} & \textbf{Patch} & \textbf{SARIF} & \textbf{Bndl} & \textbf{PoV} & \textbf{Patch} & \textbf{SARIF} & \textbf{Bndl} & \textbf{Penalty} \\
\midrule
\hyperlink{team:AT}{\texttt{AT}} & 43 & 37 & 8  & 36 & 100 & 83.8 & 75.0 & 97.2 & 0.1\% \\
\hyperlink{team:TB}{\texttt{TB}} & 31 & 24 & 1  & 19 & 90.3 & 79.2 & 100 & 100 & 0.3\% \\
\hyperlink{team:TI}{\texttt{TI}} & 55 & 63 & 7  & 19 & 61.8 & 31.7 & 62.5 & 89.5 & 7.2\% \\
\hyperlink{team:FB}{\texttt{FB}} & 35 & 60 & 7  & 13 & 80.0 & 23.3 & 87.5 & 61.5 & 8.1\% \\
\hyperlink{team:SP}{\texttt{SP}} & 31 & 11 & 12 & 7  & 90.3 & 100 & 69.2 & 100 & 0.1\% \\
\hyperlink{team:42}{\texttt{42}} & 45 & 4  & 13 & 4  & 91.1 & 75.0 & 76.9 & 100 & 0.3\% \\
\hyperlink{team:LC}{\texttt{LC}} & 1  & 3  & 0  & 1  & 100 & 33.3 & --- & 100 & 10.7\% \\
\bottomrule
\end{tabular}
}
\end{table}

%% file: tbl/fuzzing-techniques.tex
{\rowcolors{2}{}{gray!15}
\newcolumntype{L}[1]{>{\raggedright\arraybackslash}p{#1}}
\begin{table*}[t]
\centering
\caption{PoV Generation Techniques Across Teams. Blank: no custom implementation; $^\dagger$ all non-blank teams have used SARIF and Diff.
}
\label{t:fuzzing-techniques}
\smaller\smaller
\setlength{\tabcolsep}{4pt}
\adjustbox{trim=0pt 0pt 3pt 0pt,clip,max width=\textwidth}{
\begin{tabular}{@{}>{\cellcolor{white}}c L{2.8cm} L{1.59cm}L{1.59cm}L{1.59cm}L{1.59cm}L{1.59cm}L{1.59cm}L{1.59cm}@{}}
  \toprule
  & & \hyperlink{team:AT}{\texttt{AT}} & \hyperlink{team:TB}{\texttt{TB}} & \hyperlink{team:TI}{\texttt{TI}} & \hyperlink{team:FB}{\texttt{FB}} & \hyperlink{team:SP}{\texttt{SP}} & \hyperlink{team:42}{\texttt{42}} & \hyperlink{team:LC}{\texttt{LC}} \\
  \midrule
  & Pre-Comp Corpus      & \V &  & \V &  & \V & \V & \V \\
  & \tsp\tmid{} Source\newline\tsp\tcont{} & OSS-Fuzz; GitHub &  & ClusterFuzz; GitHub &  & ClusterFuzz; GitHub & OSS-Fuzz & Samples \\
  & \tsp\tlast{} Matcher & Input Format &  & Cov-Based &  & Name; Input Format & Name; Input Format & Always \\
  & LLM-Based Seed Gen & \V & \V & \V &  & \V & \V & \V \\
  & \tsp\tmid{} Bootstrap & \V & \V &  &  & \V & \V & \V \\
  & \tsp\tmid{} Solve cov blocker & Stuck Seeds & Frontier Func & Frontier Func &  & LLM-Picked &  &  \\
  & \tsp\tmid{} Mutator/generator & \V &  &  &  &  &  &  \\
  & \tsp\tmid{} Input Grammar \newline\tsp\tcont{} & Testlang; libFDP~\cite{libfdp} &  & Python Decoder &  & Nautilus~\cite{nautilus} Grammar &  &  \\
  & \tsp\tlast{} Output format & Blob; Script  & Script & Script &  & Script & Script & Blob \\
  & Engine Refinement      & \V &  &  &  & \V & \V & \V \\
  & \tsp\tmid{} Semantic feedback\newline\tsp\tcont{} &  &  &  &  & LLM for IJON~\cite{ijon} Annot. &  &  \\
  & \tsp\tmid{} Improved sanitizer & Patched (Java) &  &  &  & Loosened (Java) &  &  \\
  & \tsp\tmid{} Dict Gen\newline\tsp\tcont{}\newline\tsp\tcont{} & On-the-fly LLM &  &  &  & AFL++~\cite{AFLplusplus-Woot20} Dict2File; CodeQL~\cite{codeql} & AFL++~\cite{AFLplusplus-Woot20} Dict2File; Custom & Custom \\
  & \tsp\tlast{} Directed fuzzing & Custom distance &  &  &  &  & LLVM Slicing; WALA~\cite{wala} &  \\
  & Concolic Fuzzing       & SymCC~\cite{symcc}; Custom  &  &  &  &  &  &  \\
  & Parallel Fuzzing       & \V & \V & \V & \V & \V & \V & \V \\
  & \tsp\tmid{} Corpus sync & \V & \V & \V &  & \V & \V & \V \\
  & \tsp\tmid{} added C fuzzers\newline\tsp\tcont{}\newline\tsp\tcont{} & AFL++; libAFL~\cite{libafl}; Custom &  &  &  & AFL++ & AFL++ & AFL++ \\
  \multirow{-26}{*}[8ex]{\parbox{1.4cm}{\centering\textbf{Fuzzing Pipeline}}}
  & \tsp\tlast{} added JVM fuzzers & libAFL~\cite{libafl} &  &  &  &  &  &  \\
  \midrule
  & Bug Cand. I.D.    & \V &  & \V & \V & \V &  & \V \\
  & \tsp\tcont{} \newline \tsp\tmid{} Candidate source $^\dagger$ \newline\tsp\tcont{} \newline\tsp\tcont{} & LLM; CodeQL~\cite{codeql}; Sinks &  & LLM; Infer~\cite{infer} & LLM & LLM; Entropy; CodeQL~\cite{codeql}; Semgrep~\cite{semgrep} &  & LLM \\
  & \tsp\tcont{} \newline \tsp\tmid{} Candidate filter \newline\tsp\tcont{} & Agentic pick; Reachability &  & LLM confidence ranking & LLM pick & Multi-source weighted vote &  & Multi-LLM weighted vote \\
  & \tsp\tlast{} Non-PoV Gen usage     & \V &  & \V &  & \V &  & \V \\
  & PoV Gen Agent    & \V & \V & \V & \V & \V &  &  \\
  & \tsp\tcont{} \newline \tsp\tmid{} Key context/tool $^\dagger$\newline\tsp\tcont{} \newline\tsp\tcont{} & Code; CWE; Call Path; Cov; Log; Debugger & Code; CWE & Code; Cov; Log; Debugger & Code; CWE; Call Path; Log & Code; CWE; Call Path; Cov; Log; Debugger &  &  \\
  & \tsp\tmid{} Main method\newline\tsp\tcont{} & Iterative; Reach$\rightarrow$Exploit & Iterative & Iterative; Reach$\rightarrow$Exploit & Iterative & Iterative; Reach$\rightarrow$Exploit &  &  \\
  \multirow{-14}{*}[8ex]{\parbox{1.4cm}{\centering\textbf{LLM-Based PoV Gen Pipeline}}}
  & \tsp\tlast{} Output format     & Blob; Script & Script & Script & Script & Script &  &  \\
  \midrule
  & LLM PoV Gen $\rightarrow$ Fuzz & \V & \V & \V & \V & \V &  &  \\
  \multirow{-2}{*}{\parbox{1.4cm}{\raggedright\textbf{Pipeline Co-op}}}
  & Fuzz $\rightarrow$ LLM PoV Gen & \V &  & \V &  & \V &  &  \\
  \midrule
  & Deduplication     & Stack & ClusterFuzz~\cite{clusterfuzz} & Stack$\rightarrow$LLM classifier & Crash sig$\rightarrow$LLM & ClusterFuzz~\cite{clusterfuzz} & ClusterFuzz~\cite{clusterfuzz} & PoV hash \\
  \multirow{-3}{*}{\parbox{1.4cm}{\raggedright\textbf{PoV Submission}}}
  & Submission Strategy    & ASAP & ASAP & ASAP & ASAP & ASAP & ASAP & ASAP \\
  \bottomrule
\end{tabular}
}
\end{table*}
}

%% file: tbl/patch-techniques.tex
{\rowcolors{2}{}{gray!15}
\newcommand{\parboxSP}[1]{\parbox[c]{16em}{\centering #1}}
\newcolumntype{C}[1]{>{\centering\arraybackslash}p{#1}}
\newcolumntype{L}[1]{>{\raggedright\arraybackslash}p{#1}}
\begin{table*}[t]
\centering
\caption{Patching Techniques Used by each CRS. Blank: no custom implementation; --: not applicable; $^\dagger$all teams use sanitizer/crash reports and failed patch feedback.
}
\label{t:crs-patch-techniques}
\smaller\smaller
\setlength{\tabcolsep}{3pt}
\renewcommand{\arraystretch}{0.85}
\adjustbox{trim=0pt 0pt 3pt 0pt,clip,max width=\textwidth}{
\begin{tabular}{@{}>{\cellcolor{white}}c L{2.3cm} L{1.72cm}L{1.72cm}L{1.72cm}L{1.72cm}L{1.72cm}L{1.72cm}L{1.72cm}@{}}
  \toprule
  & & \hyperlink{team:AT}{\texttt{AT}} & \hyperlink{team:TB}{\texttt{TB}} & \hyperlink{team:TI}{\texttt{TI}} & \hyperlink{team:FB}{\texttt{FB}} & \hyperlink{team:SP}{\texttt{SP}} & \hyperlink{team:42}{\texttt{42}} & \hyperlink{team:LC}{\texttt{LC}} \\
  \midrule
  & Arch Category & Multi-Arch & Multi-Agent & Multi-Agent & Single-Agent & Multi-Arch & Single-Agent & Single-Agent \\
  & Design Detail & 6 standalone patching agents + \textsc{Aider}~\cite{aider} \par+ \textsc{SWE-Agent}~\cite{sweagent} & RCA; Strategy; Creation; Reflection & Analyzer; Patcher; Questions & Same arch with 23 strategies:\par 12 Full-mode \par 8 Delta-mode \par 2 SARIF \par 1 Unharnessed & Triage; Programmer; Critic;\par Traditional + Agentic pipelines & Test; Context; RCA; Strategy; QE & Multi-LLM workflow \\
  & Diversified Hyperparams  &  & Temp. &  &  & Temp. & Temp.; No. of failed patches & Temp. \\
  \multirow{-10}{*}[0.5ex]{\parbox{1.1cm}{\centering\textbf{Agent Architecture}}}
  & Diversified LLMs & GPT; Claude; Gemini & GPT; Claude & GPT; Claude; Gemini & Claude; GPT; Gemini & Claude; GPT & GPT; Claude; Gemini & GPT; Claude; Gemini \\
  \midrule
  & Standalone RCA & \V & \V & \V &  & \V &  &  \\
  & \tsp\tmid{} Multi-PoV \newline\tsp\tcont{} RCA\newline\tsp\tcont{} &  & Up to 15 variants & Up to 3 ranked PoVs &  & Crash statistics (AURORA~\cite{aurora}) based &  &  \\
  \multirow{-3}{*}[0ex]{\parbox{1.1cm}{\centering\textbf{Root Cause Analysis}}}
  & \tsp\tcont{}\newline\tsp\tcont{}\newline\tsp\tcont{}\newline\tsp\tlast{} Non-LLM RCA &  &  &  &  & Agent with static \& dynamic analysis;\par ensembled ranking of multi-sources &  &  \\
  \midrule
  & Contextualization$^\dagger$ & \V & \V & \V & \V & \V & \V & \V \\
  & \tsp\tmid{} Code Indexer \newline\tsp\tcont{} & ctags~\cite{ctags}; ast-grep~\cite{ast-grep} & tree-sitter~\cite{tree-sitter} & gtags~\cite{gnu-global} &  & tree-sitter~\cite{tree-sitter} & ctags; LSP~\cite{lsp} &  \\
  & \tsp\tmid{} SAST Report\newline\tsp\tcont{} &  &  & Infer~\cite{infer}; Joern~\cite{joern} & SVF~\cite{svf}; CodeQL~\cite{codeql} & Semgrep~\cite{semgrep}; CodeQL~\cite{codeql} &  &  \\
  & \tsp\tmid{} CWE Guidance\newline\tsp\tcont{} &  &  &  & 40+ CWE catalog &  & 40+ CWE repair advice &  \\
  & \tsp\tmid{} Fine-tuned\newline\tsp\tcont{} LLM \newline\tsp\tcont{} & Llama~\cite{llama} for contextualization &  &  &  &  &  &  \\
  & \tsp\tmid{} Agentic Code\newline\tsp\tcont{} Search & \V & \V & \V & \V &  \V &  \V & \V  \\
  & \tsp\tmid{} Dynamic Info \newline\tsp\tcont{} \newline\tsp\tcont{} & GDB~\cite{gdb}/JDB~\cite{jdb} &  & LLVM-cov~\cite{llvm}; JaCoCo~\cite{jacoco} & LLVM-cov/JaCoCo & GDB/JDB &  &  \\
  & \tsp\tlast{} PoV Bytes & \V &  & \V &  &  &  & \V \\
  & LLM Reflection & \V & \V & \V &  & \V &  & \V \\
  \multirow{-10}{*}[16ex]{\parbox{1.1cm}{\centering\textbf{Generation}}}
  & No-PoV Patch Generation &  &  & @45min; Delta diff; SAST & @50\%; SAST; LLM vuln ranking &  &  & Delta diff \\
  \midrule
  & Basic Checks & \V & \V & \V & \V & \V & \V & \V \\
  & \tsp\tmid{} Build & \V & \V & \V & \V & \V & \V & \V \\
  & \tsp\tmid{} PoV Test (Gen) & Single & Up to 15/san & All & Single & Up to 20/vuln & All & Single \\
  & \tsp\tlast{} Proj. Tests & \V & \V & \V & \V & \V &  & \V \\
  & PoV Test (Submit) & All \& cross-block & Up to 15/san & All & Max 5 & Up to 20/vuln & All & Single \\
  & LLM as Judge & \V &  &  & \V & \V &  &  \\
  & Post-patch Fuzz &  &  &  & 25s Fuzz (No PoV Patch Only) & 5min Fuzz &  &  \\
  \multirow{-8}{*}[4ex]{\parbox{1.1cm}{\centering\textbf{Validation}}}
  & Rebuild Optimization & ccache~\cite{ccache}; Maven~\cite{maven} cache &  &  &  & ccache; Maven cache &  &  \\
  \midrule
  & Min. Patch Set Calc. & \V & \V &  &  & \V & \V &  \\
  & \tsp\tmid{} Calc. Timing & On new PoV & On new PoV &  &  & On new PoV & Every hour &  \\
  & \tsp\tmid{} Calc. Mode \newline\tsp\tcont{} \newline\tsp\tcont{} & Incremental; Uncovered PoVs & Recompute; All PoVs &  &  & Recompute; All PoVs & Incremental; Uncovered PoVs &  \\
  & \tsp\tlast{} Submit & Right after calc.; New patch & Right after calc.; All unsubmitted &  &  & $\geq$60min; by PoV count; All unsubmitted & Right after calc.; New patch &  \\
  \multirow{-5}{*}[6ex]{\parbox{1.1cm}{\centering\textbf{De\-du\-pli\-ca\-tion \& Sub\-mis\-sion}}}
  & No-PoV Patch Sub. & -- & -- & >45min; gated by PoV success & @50\% time & -- & -- & @DDL-30min \\
  \bottomrule
\end{tabular}
}
\end{table*}
}